\documentclass[prd,showpacs,showkeys,preprintnumbers,floatfix,
nofootinbib,superscriptaddress]{revtex4}
\usepackage{float}
\usepackage{nicefrac}
\usepackage{slashed}
\usepackage{mathtools}
\usepackage{amsfonts} % AMS
\usepackage{amssymb} % AMS
\usepackage{amsmath} % AMS
\usepackage{graphicx} % Include figure files
\usepackage{subfigure} % Include figure files
\usepackage{array} % array
\usepackage{dcolumn} % Align table columns on decimal point
\usepackage{bm} % bold math
\usepackage{latexsym} % latex symbols
\usepackage{longtable} % long tables
\usepackage{hyperref} % hypertext links 
\usepackage{verbatim}
\usepackage{epsfig}
\usepackage{color}
\newcommand{\nn}[0]{\nonumber}

\begin{document}
\title{Multichannel $0 \rightarrow 2$ and $1 \rightarrow 2$ transition amplitudes\\ for arbitrary spin particles in a finite volume}
\author{Ra\'ul A. Brice\~no}
\email[e-mail: ]{rbriceno@jlab.org}
\affiliation{
Thomas Jefferson National Accelerator Facility, 12000 Jefferson Avenue, Newport News, VA 23606, USA\\
}

\author{Maxwell T. Hansen}
\email[e-mail: ]{hansen@kph.uni-mainz.de}
\affiliation{
Institut f\"ur Kernphysik and Helmholz Institute Mainz, Johannes Gutenberg-Universit\"at Mainz,
55099 Mainz, Germany\\
}
\date{\today}
\begin{abstract}

We present a model-independent, non-perturbative relation between finite-volume matrix elements and infinite-volume $\textbf{0}\rightarrow\textbf{2}$ and $\textbf{1}\rightarrow\textbf{2}$ transition amplitudes. Our result accommodates theories in which the final two-particle state is coupled to any number of other two-body channels, with all angular momentum states included.  The derivation uses generic, fully relativistic field theory, and is exact up to exponentially suppressed corrections in the lightest particle mass times the box size. This work distinguishes itself from previous studies by accommodating particles with any intrinsic spin. To illustrate the utility of our general result, we discuss how it can be implemented for studies of $N+\mathcal{J}~\rightarrow~(N\pi,N\eta,N\eta',\Sigma K,\Lambda K)$ transitions, where $\mathcal{J}$ is a generic external current. The reduction of rotational symmetry, due to the cubic finite volume, manifests in this example through the mixing of S- and P-waves when the system has nonzero total momentum. 
\end{abstract}
\pacs{12.38.Gc, 11.80.-m , 13.85.Fb, 25.20.-x} 
%12.38.Gc Lattice QCD calculations
%11.80.-m Relativistic scattering theory
%13.85.Fb Inelastic scattering: two-particle final states
%25.20.-x Photonuclear reactions

\keywords{finite volume}
\maketitle

\section{Introduction \label{sec:intro}}

Whether catalyzing the formation of exotic mesons (as is the case for the GluEx experiment~\cite{AlekSejevs:2013mkl, Shultz:2015pfa}) or probing the Standard Model of particle physics (e.g.,~heavy meson decays~\cite{Wei:2009zv,Aaltonen:2011ja,Lees:2012tva,Aaij:2013iag,Aaij:2013qta, Bobeth:2012vn,Descotes-Genon:2013wba,Hambrock:2013zya,Beaujean:2013soa, Horgan:2013hoa,Horgan:2013pva, Bouchard:2015pda, Bailey:2014fpx, Bouchard:2014ypa}), hadronic transitions lie at the core of present-day nuclear and particle physics research. Consequently, there is great incentive to provide theoretical pre- and postdictions of these reactions. However, these quantities are very challenging due to the nonperturbative nature of the strong nuclear force and its underlying quantum theory, quantum chromodynamics (QCD). The preferred approach to overcome this obstacle is to implement numerical studies of QCD, in particular lattice QCD (LQCD). 

Numerical LQCD calculations are necessarily performed in a finite Euclidean spacetime. For processes involving two or more hadrons in the initial or final states, the relationship between quantities determined from LQCD and the infinite Minkowski spacetime observables is therefore highly nontrivial.\footnote{See Refs.~\cite{Briceno:2014tqa, Briceno:2014pka, Beane:2014oea} for recent reviews on the topic.} Nonetheless much progress has been made by using finite volume as a tool, rather than an unwanted artifact, in order to extract physical observables. In particular, the problem of extracting transition amplitudes from finite-volume matrix elements was first addressed by Lellouch and L\"uscher in the context of $K\rightarrow\pi\pi$ decays~\cite{Lellouch:2000pv}. This idea has since been revisited and generalized to describe more complicated one-to-two processes~\cite{Lin:2001ek, Kim:2005gf, Christ:2005gi,  Meyer:2012wk,Hansen:2012tf, Agadjanov:2014kha, Briceno:2014uqa}, two-to-two processes~\cite{Detmold:2004qn, Briceno:2012yi, Bernard:2012bi}, as well as non-relativistic, weakly interaction multi-bosonic systems~\cite{Detmold:2014fpa}. Related ideas have also been presented to describe the finite-volume effects for the long-distance two-pion contribution to neutral kaon mixing~\cite{Christ:2010gi}.

In this work, we extend the ideas presented in our previous article, Ref.~\cite{Briceno:2014uqa}, and give a model-independent relation between finite-volume matrix elements and transition amplitudes for particles with arbitrary intrinsic spin. We primarily consider matrix elements of local currents coupling one- and two-particle finite-volume states and relate these to infinite-volume $\textbf{1}\rightarrow\textbf{2}$ transition amplitudes in the presence of an external current. We additionally consider finite-volume matrix elements connecting the vacuum to two-particle states and relate these to infinite-volume $\textbf{0}\rightarrow\textbf{2}$ transitions. Here we also present an alternative derivation from our earlier work. Instead of considering ratios of two- and three-point correlators to access the desired matrix elements, here we access them directly from a single finite-volume two-point correlator. This is possible because we have sufficient freedom for the operators appearing within the correlator and can choose these to define both $\textbf{1}\rightarrow\textbf{2}$ and $\textbf{0}\rightarrow\textbf{2}$ transitions. 

As was the case in Ref.~\cite{Briceno:2014uqa}, our result here allows for two-particle states that strongly couple to any number of open two-particle channels and also incorporates all angular momentum states. The only approximation required is neglect of exponentially suppressed corrections of the form $e^{-mL}$, where $m$ is the physical mass of the lightest particle and $L$ is the extent of the finite volume. Therefore our results apply only to theories where $mL\gg1$ and do not, for example, describe QED+QCD in finite-volume.\footnote{When performing numerical lattice QCD+QED studies, one may remove the zero mode, which allows for a systematic assessment of finite-volume effects~\cite{Duncan:1996xy, Hayakawa:2008an, Davoudi:2014qua, Borsanyi:2014jba, Carrasco:2015xwa} including for systems involving two-particles~\cite{Beane:2014qha}}
%%FOOTNOTE

Beyond facilitating the determination of a wide range of physical observables, this work will also impact the assessment of systematic error for quantities that have already been extracted from LQCD calculations. One prominent example is the study of parity violation in the nuclear sector performed by Ref.~\cite{Wasem:2011zz}. This work constrains $h^1_{\pi NN}=1.099(51)(6)\times 10^{-7}$ via LQCD using light quark masses corresponding to $m_\pi\sim389$~MeV. This is the coupling that parametrizes the parity violating $N\rightarrow \pi N$ process at threshold, where the final two-body state is in an S-wave. At the time of this benchmark calculation, it was not understood how to correct finite-volume effects due to the fact that the final state is composed of two particles. In Section~\ref{sec:NpiSPwave} we explicitly discuss the implication of the main result to $N+\mathcal{J}~\rightarrow~(N\pi,N\eta,N\eta',\Sigma K,\Lambda K)$, where $\mathcal{J}$ is any external current.

We comment that, although this work is primarily intended for application in numerical LQCD, the result presented is universal and holds for any quantum theory in a finite volume. For example, Ref.~\cite{Rupak:2013aue} studied radiative neutron capture ($n+p\rightarrow d+\gamma$) in lattice effective field theory (EFT), where nucleons are treated as fundamental degrees of freedom.%%%%FOONOTE%%%%
~\footnote{In their work, Rupak and Lee used an alternative technique to the Lellouch-L\"uscher formalism in determining the transition amplitude. }%%%%FOONOTE%%%% 
~By considering volumes where $\kappa_d L\gg 1$, where $\kappa_d$ is the binding energy of the deuteron and $L$ is the spatial extent of the volume, one may use the formalism presented to study this and many other reactions (e.g., $p+p\rightarrow d +e^++\nu_e$ or $d +\nu\rightarrow p +n+\nu$).

The remainder of this work is organized as follows. In the following section we completely specify our set-up and notation. We then define a finite-volume correlator and state a key identity, expressing it in terms of infinite-volume quantities and finite-volume kinematic functions. In Section \ref{sec:der} we review the derivation of this key identity. Then, in Section \ref{sec:QCandME}, we show how the identity for the finite-volume correlator can be used to derive all results of this work. In Appendices~\ref{sec:free_limit} and \ref{sec:narrow_width} we discuss two important simplifying limits of this result, namely the free limit and the narrow width approximation. In Appendix~\ref{sec:assym} we explain how to generalize this result to volumes that are rectangular prisms with twisted boundary conditions. 

We close this section with a summary of our main results, all derived in Section \ref{sec:QCandME} using the expression for the finite-volume correlator, Eq.~(\ref{eq:CLres}). We first show how the expression can be used to relate the finite-volume energy spectrum to two-to-two scattering amplitudes. The relation takes the form of a quantization condition~\cite{Luscher:1986pf, Luscher:1990ux, Rummukainen:1995vs, Kim:2005gf, Christ:2005gi, Briceno:2012yi, Hansen:2012tf, Briceno:2014oea} 
\begin{equation}
\label{eq:QC}
\Delta(P,L) \equiv \det[F^{-1}(P,L) + \mathcal M(P)] = 0\,,
\end{equation}
where $P$ is the four-momentum of the two scattering particles. The interpretation is that, for fixed total spatial momentum $\textbf P$ and box size $L$, the energies $E_{1,\textbf P, L}, E_{2,\textbf P, L}, \cdots$ for which $\Delta$ vanishes are precisely the energies in the spectrum of the finite-volume theory. Since $F$ is a known kinematic function [see Eq.~(\ref{eq:Fscdef})], if the energy spectrum is also known, then this result can be used to constrain the scattering amplitude, $\mathcal M$. The notation needed to completely understand this result is discussed in Section \ref{sec:CLident}. In arriving at this result, we assume the two-particle states lie below the multi-particle thresholds. For recent progress towards studying three-particle systems in a finite volume we point the reader to Refs.~\cite{Polejaeva:2012ut,Briceno:2012rv,Hansen:2014eka}.

We next derive the relation between finite-volume matrix elements and $\textbf 1 \to \textbf 2$ transition amplitudes~\cite{Lellouch:2000pv, Lin:2001ek, Kim:2005gf, Christ:2005gi, Hansen:2012tf, Agadjanov:2014kha, Briceno:2014uqa}
\begin{equation}
\vert \langle E'_n, \textbf P', L  \vert \widetilde {\mathcal J}_A(0, \textbf P - \textbf P') \vert E_{0, \textbf P}, \textbf P,L ,1\rangle \vert  = \sqrt{\frac{\mathcal{N}_{n} \mathcal{N}_1}{2E_{0,\textbf P}}} \sqrt{\mathcal H^{\mathrm{in}}_{A}  \  \mathcal R(E_{n}', \textbf P') \   \mathcal H^{\mathrm{out}}_{A} } \,,
\label{eq:matJaieps}
\end{equation}
where $\mathcal{N}_{1}$ and $\mathcal{N}_n$ are the normalization factors for the one-body and two-body finite-volume states,
\begin{align}
 \langle E_{0, \textbf P}, \textbf P,L ,1 \vert E_{0, \textbf P}, \textbf P,L ,1\rangle & = \mathcal N_1 \,, \\
 \langle E'_n, \textbf P', L  \vert E'_n, \textbf P', L  \rangle & = \mathcal N_n \,.
\end{align}
We set these equal to one for the remainder of this work. Here we have also introduced
\begin{equation}
\label{eq:Rintro}
\mathcal R(E_{n}, \textbf P) \equiv  \lim_{P_4 \rightarrow i E_{n}} \left[ - (i P_4 + E_{n}) \frac{1}{F^{-1}(P,L) + \mathcal M(P)}\right] \,,
\end{equation}
\begin{equation}
\label{eq:HinIntro}
\big[ \mathcal H^{\mathrm{out}}_{A}(E_{0,\textbf P}, \textbf P; E_{n}', \textbf P') \big] \ (2 \pi)^3  \delta^3(\textbf P - \textbf P' - \textbf Q) \equiv  \langle E'_{n}, \textbf P',\{J'\},\mathrm{out} \vert \widetilde {\mathcal J}_A(0, \textbf Q)\vert E_{0,\textbf P} , \textbf P,\{J\}_1   \rangle \,.
\end{equation}
Here and below we use $E_n$ in place of $E_{n, \textbf P, L}$ to reduce clutter of notation. In Eq.~(\ref{eq:HinIntro}) we have introduced the labels $\{J\}_1$ and $\{J'\}$, which stand for lists of all quantum numbers specifying the one-particle in-state and two-particle out-state respectively. For the out-state, $\{J'\}$ includes a channel label denoting the flavor of the individual outgoing particles. In Eq.~(\ref{eq:matJaieps}), $\mathcal H^{\mathrm{in}}$ is understood as a row, $\mathcal R$ as a matrix and $\mathcal H ^{\mathrm{out}}$ as a column in this index space. $\mathcal J_A$ is a generic external local current. The matrix $\mathcal R$ is the generalization of the Lellouch-L\"uscher proportionality factor that enters the relation for extracting $K \rightarrow \pi \pi$ decays. Note that evaluating the limit in practice requires determining the first-derivative of the scattering amplitude evaluated at the energy pole. Again, see Section \ref{sec:CLident} for a complete explanation of notation.

Finally, Section \ref{sec:QCandME} contains the derivation of two additional results. First we present a method for constraining the relative sign of two different transition amplitudes, involving the same in- and out-states but different current insertions 
\begin{equation}
\frac{\langle E'_n, \textbf P', L  \vert \widetilde {\mathcal J}_{A_1}(0, \textbf P - \textbf P') \vert E_{0, \textbf P}, \textbf P,L ,1\rangle}{\langle E'_n, \textbf P', L  \vert \widetilde {\mathcal J}_{A_2}(0, \textbf P - \textbf P') \vert E_{0, \textbf P}, \textbf P,L ,1\rangle}   = \frac{\mathcal X^\dagger \  \mathcal R(E_{n}', \textbf P') \  \mathcal H^{\mathrm{out}}_{A_1} }{\mathcal X^\dagger \ \mathcal R(E_{n}', \textbf P')\  \mathcal H^{\mathrm{out}}_{A_2} }\,,
\label{eq:ratio}
\end{equation}
where $\mathcal X^\dagger$ is a completely arbitrary vector to be chosen at the user's convenience. This result follows from the observation that $\mathcal R$ can be written as an outer product of row and column vectors [see Eq.~(\ref{eq:outerproduct})] below. This outer-product form suggests an interpretation of Eqs.~(\ref{eq:matJaieps}) and (\ref{eq:ratio}) as relations between finite- and infinite-volume two-particle states, an idea already introduced in Ref.~\cite{Lellouch:2000pv} and also discussed in Refs.~\cite{Meyer:2012wk,Hansen:2012tf}. 

Second, we derive the results for $\textbf{0} \rightarrow \textbf{2}$ transitions, which are reached by a straightforward modification of those just given. We find
\begin{equation}
\label{eq:matJ0to1}
\vert \langle E_n, \textbf P, L \vert \widetilde {\mathcal J}_A(0, -\textbf P) \vert 0 \rangle \vert  = \sqrt{\mathcal{N}_{n}}~\sqrt{\mathcal V^{\mathrm{in}}_{A}  \Big[ L^3 \mathcal R(E_{n}, \textbf P)\Big ]  \mathcal V_{A}^{\rm out} } \,, 
\end{equation}
where
\begin{equation}
\big [ \mathcal V_{A}^{\rm out}(E_{n}, \textbf P) \big ]\ (2 \pi)^3  \delta^3(\textbf P + \textbf Q) \equiv \langle E_{n}, \textbf P,\mathrm{out} \vert \widetilde {\mathcal J}_A(0, \textbf Q)\vert 0  \rangle \,.
\end{equation}
We emphasize that the formalism presented here only holds for local currents and cannot be implemented for processes that contain long range contributions, e.g., $\gamma^*\gamma^*\rightarrow\pi\pi$.
 
This equation will allow for future studies of decay constants of hadronic resonances. One example that has received attention in the lattice QCD community is the decay constant of the $\pi(1300)$ resonance~\cite{McNeile:2006qy, Mastropas:2014fsa}. To be able to study this resonance at or near the physical point would require solving the many-body, multi-coupled-channel problem in a finite volume. This is beyond the scope of current calculations due to the issues discussed above. For unphysically heavy light quark masses, however, the possible strong decay channels reduce dramatically. For instance, at the SU(3) flavor point, studied in Ref.~\cite{Mastropas:2014fsa}, the only open channel is the $\rho\pi$-scattering channel, where the $\rho$ is itself stable under the strong interactions. In this limit, one can utilize Eq.~(\ref{eq:matJ0to1}) to determine the $\rho\pi\rightarrow\textbf{0}$ transition amplitude. From this, one may analytically continue the $\pi(1300)$ pole, and thereby determine its decay constant.

 \section{Identity for the finite-volume correlator}

\label{sec:CLident}

In this section we review the decomposition of the two-to-two finite-volume correlator, denoted $C_L(P)$, into products of finite- and infinite-volume quantities. This decomposition was presented for scalar particles in Ref.~\cite{Kim:2005gf} and for the $N \pi$ system in the $\Delta$-resonance channel in Refs.~\cite{Li:2012bi, Agadjanov:2014kha}. The result reviewed here accommodates particles with arbitrary spin. We begin by describing the set-up of the calculation as well as the notation that we use.

In this work we describe particles and interactions using a completely general relativistic field theory. [See Eq.~(\ref{eq:lagdef}) below and the discussion that follows.] To restrict attention to two-particle states we require that the center-of-mass (CM) frame energy is below the lowest threshold with more than two particles. Thus, all on-shell states contain exactly two particles. Here we accommodate $N_c$ different channels each containing a particle pair with given individual physical masses and spins. We use the index $a=1,\cdots,N_c$ to label the channels considered; $m_{a1}$, $m_{a2}$ denote the two physical (pole) masses and $s_1^a$, $s_2^a$ denote the two spins.

This theory is then considered in a finite-volume. In the present work we restrict attention to the simplest possible implementation, a finite cubic spatial volume with linear extent, $L$, and with periodic boundary conditions. The generalization to asymmetric boxes and to twisted boundary conditions is well understood and we give the equations in Appendix \ref{sec:assym}. We further assume that the time extent is taken to be infinite, and that $L$ is large enough such that exponentially suppressed corrections of the form $e^{-mL}$ can be neglected, with $m$ the physical mass of the lightest particle in the spectrum. In our analysis we allow for nonzero total momentum in the finite-volume frame. We use $\textbf P$ to denote the total momentum in this frame and $E$ to denote the total energy. Due to the periodic boundary conditions, $\textbf P$ must equal a vector of integers multiplied by $(2 \pi/L)$ [$\textbf P \in (2 \pi/L) \mathbb Z^3$]. Finally we introduce $E^* = \sqrt{E^2 - \textbf P^2}$ for the CM frame energy. As already mentioned, $E^*$ is constrained to a range of energies for which only two-particle states can go on-shell.\footnote{In the case of a $\mathbb Z_2$ symmetry that decouples even- and odd-particle-number states, the required restriction is $0<E^*<4m$.} 

Within this finite-volume set-up we define the correlator
\begin{equation}
C_L(P) = C_L(P_4, \textbf P) \equiv \int_L d^4 x e^{- i P x} \Big [ \langle 0 \vert T \mathcal A(x) \mathcal B^\dagger(0) \vert 0 \rangle \Big ]_L \,.
\end{equation}
Here $T$ indicates time ordering and $\mathcal A(x)$, $\mathcal B^\dagger(x)$ are operators which couple to the two-particle states of interest. The general allowed forms for $\mathcal A(x)$ and $\mathcal B^\dagger(x)$ are of great importance to the arguments made in Section \ref{sec:QCandME}. We discuss the general conditions for which our analysis of $C_L$ holds in the text following Eq.~(\ref{eq:ALzerodef}) below and again in Section \ref{sec:QCandME} where specific choices for the operators are considered. We use the Euclidean signature metric throughout with $x_4 \equiv i x^0$ and $P_4 = i P^0 = i E$. The subscript $L$ on the correlator indicates that it is to be evaluated using the Feymann rules of finite-volume field theory. As was shown in Ref.~\cite{Kim:2005gf}, one can express $C_L(P)$ in terms of the infinite-volume correlator $C_\infty(P)$, infinite-volume matrix elements of $\mathcal A(0)$, $\mathcal B^\dagger(0)$, infinite-volume two-to-two scattering amplitudes, and known finite-volume kinematic functions. We carefully define the matrix elements, scattering amplitudes and kinematic functions and then state the result for $C_L(P)$ in Eq.~(\ref{eq:CLres}) below.

To define the infinite-volume matrix elements of $\mathcal A(0)$ and $\mathcal B^\dagger(0)$, as well as infinite-volume scattering amplitudes, we first need to introduce notation for two-particle states with arbitrary spin. We first consider two-scalar states. These may be specified by total energy $E$, total momentum $\textbf P$, momentum of a single particle $\textbf k$ and channel $a$: $\vert E , \textbf P, a, \textbf k \rangle$. The states are assumed to satisfy standard relativistic normalization
\begin{equation}
\label{eq:twopartnorm}
\langle E' , \textbf P', a', \textbf k' \vert E , \textbf P, a, \textbf k \rangle =\delta_{aa'} 2 \omega_{1a} 2 \omega_{2a} \left[ \delta^3(\textbf k - \textbf k') \delta^3(\textbf P - \textbf k - \textbf P' + \textbf k') + \delta(a) \delta^3(\textbf k - \textbf P' + \textbf k') \delta^3(\textbf P - \textbf k - \textbf k') \right] \,,
\end{equation}
where $\delta(a)=1$ if the particles are identical and $0$ otherwise. We have also introduced the shorthand
\begin{equation}
\omega_{a1} = \sqrt{m^2_{a1} + \textbf k^2} \,, \ \ \ \ \omega_{a2} = \sqrt{m^2_{a2} + (\textbf P - \textbf k)^2} \,.
\end{equation}

It is important to note that, with $E$ and $\textbf P$ fixed, $\textbf k$ cannot be freely chosen. Indeed the only remaining degree of freedom for two on-shell scalars is the direction of motion for one of the two in the CM frame. To see this, note that if $\textbf k$ is chosen to satisfy the total energy condition $E=\omega_{a1}+\omega_{a2}$, then a boost of the four-vectors $(\omega_{a1}, \textbf k)$ and $(\omega_{a2}, \textbf P - \textbf k)$ to the CM frame, i.e.~with boost velocity $\boldsymbol \beta = - \textbf P/E$, gives the four-vectors $(\sqrt{m_{a1}^2 + q^{*2}_a} , q^*_a \hat {\textbf k}^*_a)$ and $(\sqrt{m_{a2}^2 + q^{*2}_a} , - q^*_a \hat {\textbf k}^*_a)$ respectively. Here $q^*_a$, the magnitude of particle momentum in the CM frame, is given by
\begin{equation}
E^* = \sqrt{m_{a1}^2 + q^{*2}_a} + \sqrt{m_{a2}^2 + q^{*2}_a}  \,.
\end{equation}
We deduce that $\hat {\textbf k}^*_a$ is the only remaining degree of freedom for fixed $E, \textbf P$ so that we can rewrite the two-scalar state as $\vert E, \textbf P, a,\hat{\textbf k}^*_a \rangle$. It is further convenient to decompose the two-particle states in spherical harmonics. In this work we will need to consider both in- and out-states and we thus define
\begin{align}
\vert E, \textbf P, a,\hat{\textbf k}^*_a ,\mathrm{in} \rangle & = \sqrt{4 \pi} Y_{l m_l}(\hat{\textbf k}^*_a) \vert E, \textbf P, a,l ,m_l,\mathrm{in} \rangle \,, \\
\langle E, \textbf P, a',\hat{\textbf k}^*_{a'} ,\mathrm{out} \vert & = \sqrt{4 \pi} Y^*_{l' m_l'}(\hat{\textbf k}^*_{a'}) \langle E, \textbf P, a',l' ,m_l',\mathrm{out} \vert \,.
\end{align}

To include spin we simply add the labels $s_1^a, m_{s_1}, s_2^a, m_{s_2}$ to the two-particle states. Here $m_{s_j}$ labels the azimuthal component of the spin of the $jth$ particle, with respect to some arbitrary fixed axis in that individual particle's CM frame. A particularly convenient choice is to quantize the spin of the particle along its finite-volume-frame momentum. This defines the helicity, which we label as $\lambda_{s_j}$. Since the full set of indices is long we also introduce the shorthand
\begin{align}
\vert E, \textbf P, \{ l\}, \mathrm{in} \rangle & \equiv \vert E, \textbf P, a,l ,m_l,s_1^a, \lambda_{s_1}, s_2^a, \lambda_{s_2},\mathrm{in} \rangle\,, \\
\langle E, \textbf P, \{ l' \}, \mathrm{out} \vert & \equiv \langle E, \textbf P, a',l' ,m_l',s'^a_1, \lambda_{s_1}', s'^a_2, \lambda_{s_2}',\mathrm{out} \vert\,.
\end{align}
An alternative basis for two particles with spin is reached by replacing the indices $l ,m_l,s_1^a, \lambda_{s_1}, s_2^a, \lambda_{s_2}$ with $J, M, l, S, s_1^a,s_2^a$. Here $J$ is total angular-momentum in the CM frame, $M$ is the azimuthal component of total-angular momentum in the CM frame and $S$ is the total spin. The two bases are related by
\begin{equation}
\label{eq:ltoJtrans}
\vert E, \textbf P, \{J\},\mathrm{in} \rangle= \sum_{\lambda_{s_1}, \lambda_{s_2}, m_{l}, m_S} \vert E, \textbf P, \{l\},\mathrm{in} \rangle \langle \{l\} \vert \{J\} \rangle \,,
\end{equation}
where we have introduced analogous shorthand for the other basis
\begin{align}
\vert E, \textbf P, \{ J\}, \mathrm{in} \rangle & \equiv \vert E, \textbf P, a,J, M, l, S, s_1^a,s_2^a,\mathrm{in} \rangle\,, \\
\langle E, \textbf P, \{ J' \}, \mathrm{out} \vert & \equiv \langle E, \textbf P, a',J', M', l', S', s'^a_1,s'^a_2,\mathrm{out} \vert\,,
\end{align}
and where
\begin{equation}
\label{eq:CG}
\langle \{l\} \vert \{J\} \rangle = \langle s_1^a \, \lambda_{s_1}, s_2^a\, \lambda_{s_2} \vert S \, m_S \rangle \langle l \, m_l, S\, m_S \vert J \, m_J \rangle \,,
\end{equation}
with the right-hand side equal to a product of helicity Clebsch-Gordan coefficients. Note that the Clebsch-Gordan coefficients are more commonly written in terms of states whose spin is quantized along the $\hat{z}$-axis, $\langle s_1^a \,m_{s_1}, s_2^a\, m_{s_2} \vert S \, m_S \rangle$. One can write $\langle \{l\} \vert \{J\} \rangle$ in terms of these and Wigner-$\mathcal{D}$ matrices. Let $R_j$ be an active rotation from $(0,0,|\textbf{k}_j|)$ to $\textbf{k}_j$, where $\textbf{k}_j$ is the momentum of the $jth$ particle. Then define $\mathcal{D}^{({s_j})}_{m_{s_j}\lambda_j}(R_j)$ as the $m_{s_j}\lambda_j$ component of the corresponding Wigner-$\mathcal{D}$ matrix in the $s_j$ representation. With this, we find that Eq.~(\ref{eq:CG}) can be written as
\begin{align}
\label{eq:wigD}
\langle \{l\} \vert \{J\} \rangle = \sum_{m_{s_1},m_{s_2}}
\langle s_1^a \, m_{s_1}, s_2^a\, m_{s_2} \vert S \, m_S \rangle \langle l \, m_l, S\, m_S \vert J \, m_J \rangle
\mathcal{D}^{({s_1})*}_{m_{s_1}\lambda_1}(R_1)
\mathcal{D}^{({s_2})*}_{m_{s_2}\lambda_2}(R_2)
 \,.
\end{align}

Having completed our introduction of two-particle states, we now define the infinite-volume matrix elements of $\mathcal A(0)$ and $\mathcal B^\dagger(0)$ which enter the identity for $C_L(P)$ 
\begin{align}
\label{eq:Aldef}
A_{\{l\}}(P) & \equiv  A_{\{l\}}(P_4,\textbf P) \equiv \langle 0 \vert  \mathcal A(0) \vert \!-\!iP_4,  \textbf P,\{l\}, \mathrm{in} \rangle \,, \\
B^{\dagger}_{\{l'\}}(P) & \equiv B^{\dagger}_{\{l'\}}(P_4, \textbf P) \equiv \langle -iP_4, \textbf P, \{l'\}, \mathrm{out} \vert  \mathcal B^\dagger(0) \vert 0 \rangle \,,
\end{align}
or in the total-angular-momentum basis
\begin{align}
A_{\{J\}}(P) & \equiv  A_{\{J\}}(P_4,\textbf P) \equiv \langle 0 \vert  \mathcal A(0) \vert \!-\!iP_4,  \textbf P,\{J\}, \mathrm{in} \rangle \,, \\
B^{\dagger}_{\{J'\}}(P) & \equiv B^{\dagger}_{\{J'\}}(P_4 , \textbf P) \equiv \langle - i P_4, \textbf P, \{J'\}, \mathrm{out} \vert  \mathcal B^\dagger(0) \vert 0 \rangle \,.
\label{eq:BJdef}
\end{align}
A subtlety arises at this stage. The states appearing in the equations above can only be interpreted as two-particle states for $-i P_4 = E$ real and positive. However the field theoretic definition of the states can be extended into the complex plane so that $A(P)$ and $B^{\dagger}(P)$ can be defined for all values of $P_4$. We assume that this extension has been performed but it will play little role and will not enter our final result. Finally, below we will often suppress the indices and write $A(P)$ to represent a row-vector and $B^{\dagger}(P)$ to represent a column. In all expressions where this is done the equation is correct in both $\{l\}$ and $\{J\}$ bases.

We turn next to the two-to-two scattering amplitudes, collectively denoted $\mathcal{M}$, which enter the identity for $C_L(P)$. These are given by  
\begin{align} 
\big [ \mathcal M_{\{l'\}, \{l\}}(P) \big ]\, (2 \pi)^4 \delta^4(P-P') & \equiv \langle -iP_4' , \textbf P',\{l'\}, \mathrm{out}  \vert \!-\!iP_4 , \textbf P, \{l\}, \mathrm{in} \rangle_{\mathrm{conn}} 
\,, \\
\big [\mathcal M_{\{J'\}, \{J\}}(P) \big ] \, (2 \pi)^4 \delta^4(P -  P') & \equiv \langle -iP_4' , \textbf P',\{J'\}, \mathrm{out}  \vert \!-\!iP_4 , \textbf P, \{J\}, \mathrm{in} \rangle_{\mathrm{conn}} 
\,,
\end{align}
where the subscript ``conn'' indicates that only fully connected diagrams should be included. These scattering amplitudes are related to the S-matrix via
\begin{equation}
\label{Smatrix}
i\mathcal{M}(P) \equiv \mathbb{P}^{-1} {\big [S(P)-\mathbb{I} \big ]} \mathbb{P}^{-1} \,.
\end{equation}
Here $\mathbb I$ is the identity element, corresponding to the disconnected diagrams that we omit, and
\begin{equation}
\mathbb{P}=\frac{1}{\sqrt{4\pi E^*}} \text{diag} \left (\sqrt{\xi_1q^{*}_1},\sqrt{\xi_2q^{*}_2},\ldots,\sqrt{\xi_Nq^{*}_N} \right ) \,,
\end{equation}
with $\xi_a$ equal to 1/2 if the particles in the $a$th channel are indistinguishable and equal to 1 otherwise. For $-i P_4=E$ real and positive, this is the standard definition of the two-to-two scattering amplitude and S-matrix. Analytic continuation to the entire complex $P_4$ plane is also well understood.  As with $A(P)$ and $B^{\dagger}(P)$, we will often suppress the indices below and write $\mathcal M(P)$ which is understood as a matrix.

Finally, the kinematic quantities which enter the identity for $C_L(P)$ are
\begin{align}
\label{eq:Flldef}
F_{\{l\},\{l'\}}(P,L)  & \equiv   \delta_{\lambda_{s_1} \lambda_{s_1}'} \delta_{\lambda_{s_2} \lambda_{s_2}'}  F^{\mathrm{sc}}_{alm,a'l'm'}(P,L)    \,, \\
\label{eq:FJJdef}
F_{\{J\},\{J'\}}(P,L) & \equiv  \delta_{SS'}  \sum_{m_l,m'_l,m_S} \langle l \,m_l, S \, m_S \vert J M \rangle  \langle l' \,m_l', S' \, m_S \vert J' M' \rangle  F^{\mathrm{sc}}_{alm,a'l'm'}(P,L)    \,, 
\end{align}
where
\begin{equation}
\label{eq:Fscdef}
F^{\mathrm{sc}}_{alm_l,a'l'm'_l}(P,L)  \equiv \xi_a \delta_{aa'} 
\left[\frac{1}{L^3}\sum_{\mathbf{k}}\hspace{-.5cm}\int~\right]
\frac{ 4 \pi  Y_{lm_l}(\hat {\textbf k}^*_a)Y^*_{l'm'_l}(\hat {\textbf k}^*_a)  }{2 \omega_{a 1} 2 \omega_{a 2}(E -  \omega_{a 1} - \omega_{a 2} + i \epsilon )} \left (\frac{k^{*}_a}{q^*_a} \right)^{l+l'} \,.
\end{equation}
Here the superscript ``sc'' stands for scalar. We have also introduced the notation
\begin{equation}
\label{eq:sumint_diff}
 \left[ \frac{1}{L^3}\sum_{\mathbf{k}}\hspace{-.5cm}\int~\right]\equiv\bigg [ \frac{1}{L^3} \sum_{\textbf k \in (2 \pi/L) \mathbb Z^3} - \int \frac{d\textbf k}{(2 \pi)^3} \bigg ] \,.
\end{equation}
In Eq.~(\ref{eq:Fscdef}), $\textbf k^*_a = k^*_a \hat{\textbf k}_a^*$ is the spatial component of $(\omega_{a1}^*, \textbf k^*_a)$, which is the four vector reached by boosting $(\omega_{a1}, \textbf k)$ with a boost velocity $\boldsymbol \beta = - \textbf P/E$. Note that the magnitude $k^*_a$ is not equal generally equal to $q^*_a$ introduced above, because, for a general choice of $\textbf k$, the condition $E = \omega_{a1}+\omega_{a2}$ is not satisfied. See Eq.~(\ref{eq:FKSS}) below for the alternative definition of $F^{\mathrm{sc}}$ that was first given in Ref.~\cite{Kim:2005gf}. 

Our claim, embodied in Eq.~(\ref{eq:Flldef}), is that spin can be incorporated by trivially modifying the results for scalar particles derived in Ref.~\cite{Kim:2005gf}. We demonstrate in the next section that this straightforward modification, simply including Kronecker deltas in helicity, is the correct prescription. Using the total angular momentum basis, Eq.~(\ref{eq:FJJdef}), results in a more significant difference in the form of $F$ between scalar and non-scalar particles. The relations between the two basis follow from applying the transformations specified in Eqs.~(\ref{eq:ltoJtrans}) - (\ref{eq:wigD}). Note that the distinction between helicity quantization and spin quantization along some arbitrary axis does not enter into Eq.~(\ref{eq:FJJdef}). In other words, the Wigner-D matrices appearing in Eq.~(\ref{eq:wigD}) cancel out in the definition of $F_{\{J\},\{J'\}}$.

We are now ready to state the key identity for $C_L(P)$
\begin{equation}
\label{eq:CLres}
C_L(P) = C_{\infty}(P) - A(P) \frac{1}{F^{-1}(P,L) + \mathcal M(P)} B^{\dagger}(P) \,.
\end{equation}
Since $A(P)$ is a row-vector, $\mathcal M(P)$ and $F(P,L)$ are matrices, and $B^{
\dagger}(P)$ is a column vector, the second term of Eq.~(\ref{eq:CLres}) has no uncontracted indices. It is for this reason that the second term is basis independent, and can be expressed using both $\{l\}$ and $\{J\}$ indices. In the following section we review the derivation of this result, and in Section \ref{sec:QCandME} we use it to derive our relations for extracting $\textbf{0}\rightarrow\textbf{2}$ and $\textbf{1}\rightarrow\textbf{2}$ transition amplitudes. For the latter case, we reach the relation by choosing $\mathcal A$ and $\mathcal B^\dagger$ to contain both the current and the single-particle creation operators. It is for this reason that we must also carefully consider the types of operators $\mathcal A$ and $\mathcal B^\dagger$ for which Eq.~(\ref{eq:CLres}) holds, to justify this non-standard construction.

%%%%%%%%%%%%%%%%%%%%%%%%%%%%%%%%%%%%%%%%%%%
\begin{figure*}[t]
\begin{center}
\subfigure[]{
\label{fig:FVcorr}
\includegraphics[scale=0.45]{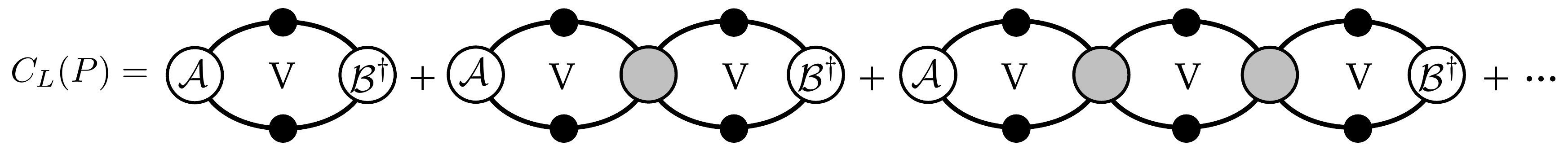}}\\
\subfigure[]{
\label{fig:kernel}
\includegraphics[scale=0.25]{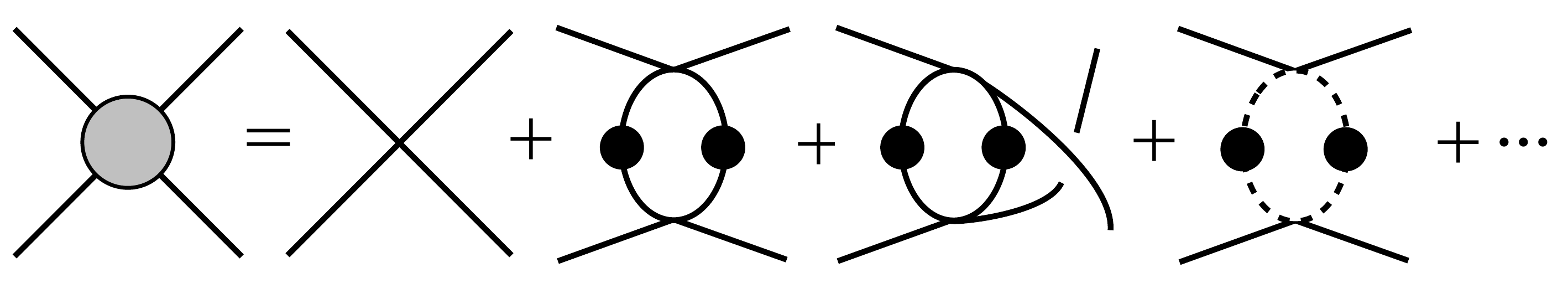}}
\subfigure[]{
\label{fig:1bodyprop}
\includegraphics[scale=0.25]{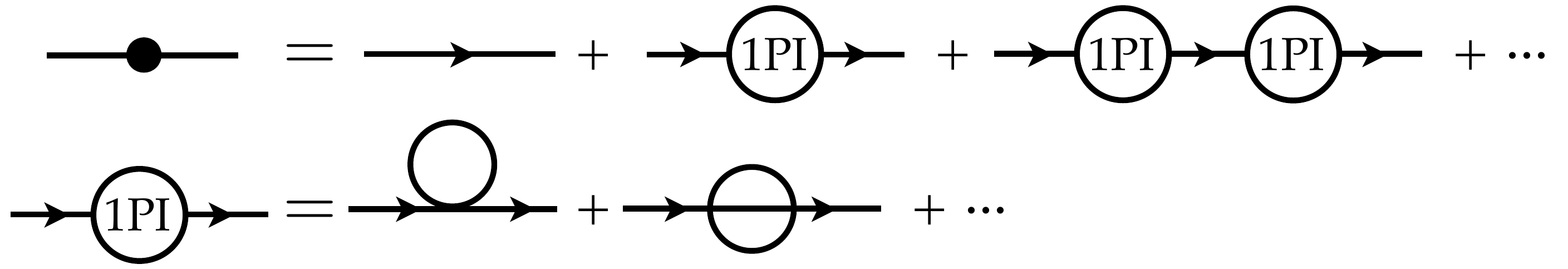}}

\caption{(a) Depicted is the diagrammatic representation of the two-point correlation function in a finite volume for energies where only two-particle states can go on-shell.  $\mathcal B^\dagger$ and $\mathcal A$ denote the creation and annihilation operators respectively. The gray circles depict the sum of all diagrams that are two-particle irreducible in the $(E, \textbf P)$-channel, defined in (b). The propagators are fully dressed and explicitly defined in terms of the one-particle irreducible (1PI) diagrams in  (c). Although in the figure we only explicitly depict the contributions to the correlation function of a single channel, the formalism presented holds for an arbitrary number of open, two-body channels. See Ref.~\cite{Briceno:2014uqa} for a diagrammatic representation with an arbitrary number of open channels. }\label{fig:corr2}
\end{center}
\end{figure*}
%%%%%%%%%%%%%%%%%%%%%%%%%%%%%%%%%%%%%%%%%%%

\section{Derivation of Eq.~(\ref{eq:CLres})}

\label{sec:der}

In this section we review the derivation of Eq.~(\ref{eq:CLres}) in three parts:
\begin{enumerate}
\item Argue that $C_L(P)$ is equal to an infinite diagrammatic series which can be organized into a skeleton expansion built from two-particle loops (with summed momenta) and infinite-volume two-to-two Bethe-Salpeter kernels [see Figure~\ref{fig:corr2}].
\item Apply an identity which rewrites the summed two-particle loops as analogous integrated loops plus a finite-volume residue correction, given by $F(P,L)$ [see Figure~\ref{fig:F_function}].
\item Reorganize the terms in the series by the number of factors of $F$ they contain, and then sum all infinite-volume diagrams which appear outside and between $F$ factors. We find that the infinite-volume diagrams outside $F$ factors sum to $A(P)$, $B^{\dagger}(P)$ and those between to $\mathcal M(P)$. Summing the series to all orders in $F$ gives Eq.~(\ref{eq:CLres}) [see Figure~\ref{fig:CV_m_Cinfy}].
\end{enumerate}
We now discuss each of the steps in some detail.

\subsection{Skeleton expansion}

To reach the skeleton expansion we start with the general diagrammatic expansion which defines $C_L(P)$. This is determined by a set of Feynman rules which are in turn specified by the Lagrangian density of the theory
\begin{equation}
\label{eq:lagdef}
\mathcal L = (1/2) \sum_{i=1}^{N_s} \phi_i \big ( \partial^2 +m_{s,i}^2 \big ) \phi_i +  \sum_{j=1}^{N_f} \overline \Psi \big (\slashed \partial + m_{f,j}  \big ) \Psi + \cdots + V(\phi, \Psi, \cdots) \,.
\end{equation}
Note that all terms are written with Euclidean signature, for example $\partial^2 \equiv \sum_\mu \partial_\mu \partial_\mu$. Here we have explicitly shown $N_s$ different scalar particles, with corresponding fields $\phi_1, \cdots, \phi_{N_s}$, as well as $N_f$ different spin-1/2 fermions, denoted $\Psi_1, \cdots, \Psi_{N_f}$. To define the fermion kinetic terms we have introduced the shorthand $\slashed \partial \equiv \partial_\mu \gamma_{E,\mu}$ where $\gamma_{E,\mu}$ are Euclidean gamma matrices defined to satisfy $\{ \gamma_{E,\mu}, \gamma_{E,\nu}\} = 2 \delta_{\mu \nu} $. The first ellipsis in Eq.~(\ref{eq:lagdef}) stands for the kinetic terms of other particle types, with spin exceeding one half. All possible interactions are include in the potential $V(\phi, \Psi, \cdots)$, where the ellipsis stands for the dependence on all other particle types. These could include fields for spin-1 mesons (e.g., the $J/\Psi$ and $B^*$), stable spin-3/2 baryons (e.g. the $\Omega$-baryon), etc. The interactions are assumed to be local and to respect Lorentz invariance but are otherwise arbitrary. We further assume that all counter terms, including mass and wave-function renormalization, are included in $V(\phi, \Psi, \cdots)$. We constrain these using on-shell renormalization, meaning that all $m$ parameters in the Lagrangian density are physical pole masses and that the wave-function renormalization of all fields is equal to one. We stress here that this Lagrangian is understood to be in terms of the low-energy degrees of freedom of the theory, so that there exists a correspondence between the fields shown and the low-lying particle spectrum.

The resulting Feynman rules include standard propagators of the form
\begin{equation}
\Delta^0_{i,\mathrm{free}}(p) \equiv \frac{1}{p^2 + m_{s, i}^2} \,,  \ \  \Delta^{1/2}_{j,\mathrm{free}}(p) \equiv \frac{- i \slashed p + m_{ f, j}}{p^2 + m_{f, j}^2} \,, \ \ \cdots \,,
\end{equation}
where, as above, the ellipsis stands for the analogous propagators of higher spin fields. We will return to the specific forms of the higher spin propagators below. Interactions and counterterms are encoded by a set of vertices, whose specific form is determined by $V(\phi, \Psi, \cdots)$. 

To calculate $C_L(P)$, one also requires the Feynman rules induced by the interpolators $\mathcal A$ and $\mathcal B^\dagger$. These are determined by first expressing the operators in terms of the elementary fields appearing in Eq.~(\ref{eq:lagdef}). For scalar particles, for example, the decompositions generically take the form
\begin{equation}
\label{eq:Aopdef}
\mathcal A(x) \equiv \sum_{i} \int \! d^4 y \, A^{(0)}_i(y) \phi_i(x+y) +  \sum_{ij} \xi_{ij} \int \! d^4 y \, \int \! d^4 z \, A^{(0)}_{ij}(y,z) \phi_{i}(x+y)  \phi_{j}(x+z) + \cdots \,,
\end{equation}
and similar for $\mathcal B^\dagger$. Here the ellipsis stands for terms constructed from three or more single particle fields. In each term the operators are integrated over weight functions, denoted $A^{(0)}$. Note that the weight functions cannot depend on $x$ as this would violate the translation property
\begin{equation}
e^{-i \hat {\textbf P} \cdot \textbf y + \hat H y_4} \mathcal A(x) e^{i \hat {\textbf P} \cdot \textbf y - \hat H y_4} = \mathcal A(x+y) \,,
\end{equation}
which must hold if $C_L(P)$ is to be projected to a single total momentum. The indices $ij \cdots$ are applied to the weight functions since these can be different for different particle flavors. These are also applied to the symmetry factors $\xi_{ij \cdots}$, which equal prodcuts of $1/n!$ factors for each subset of $n$ identical scalars. 

Turning next to spin-one-half particles, we generalize the operator decomposition by replacing the scalar fields with Dirac fields and then additionally including Dirac indices on the weight functions
\begin{equation}
\label{eq:Aopdefspin}
\mathcal A(x) \equiv \sum_{i} \int \! d^4 y \, A^{(0)}_{i,\alpha_i}(y) \Psi_{i,\alpha_i}(x+y) +  \sum_{ij} \xi_{ij} \int \! d^4 y \, \int \! d^4 z \, A^{(0)}_{ij,\alpha_i,\alpha_j}(y,z) \Psi_{i,\alpha_i}(x+y)  \Psi_{j,\alpha_j}(x+z) + \cdots \,.
\end{equation}
This form readily generalizes to also accommodate higher spin particles. For a particle with spin $s$ there must exist some ``Dirac-like'' indices which encode the spin degrees of freedom. These indices generally do not correspond to definite spin states. In the following section we will describe a change of basis that converts such {\em field indices}, which label the interpolating functions and propagators, to spin indices that label physical states. However, as a first step we express the operators in terms of indexed weight functions as in Eq.~(\ref{eq:Aopdefspin}). We finally note that, for our fully general theory, the operators $\mathcal A$ and $\mathcal B^\dagger$ will be a sum over all of the different spin types included in the theory.

Having discussed the general forms of the interpolators $\mathcal A$ and $\mathcal B^\dagger$ in terms of single-particle fields, we are now in position to give the Feynman rules associated with these operators. These are given by Fourier transforming the weight functions of Eqs.~(\ref{eq:Aopdef}) and (\ref{eq:Aopdefspin}). For example the single field terms of $\mathcal A$ and $\mathcal B^\dagger$ in the scalar sector, first term in Eq.~(\ref{eq:Aopdef}), induce single-legged vertices with vertex factors
\begin{align}
\label{eq:Aoneleg}
\tilde A^{(0)}_i & = \int \! d^4 y \, e^{i P y} A^{(0)}_i(y) \,, \\
\tilde B^{(0) \dagger}_i & = \int \! d^4 y \, e^{-i P y} B^{(0) \dagger}_i(y) \,,
\end{align}
where here and below we suppress the dependence on total momentum $P$. Similarly, the two-field terms induce two legged vertices with factors
\begin{align}
\tilde A^{(0)}_{ij}(k) & = \int \! d^4 y \! \int \! d^4 z \, e^{i (P-k) y} e^{ikz} A^{(0)}_{ij}(y,z) \,, \\
\tilde B^{(0) \dagger}_{ij}(k) & = \int \! d^4 y \! \int \! d^4 z \,  e^{-i (P-k) y} e^{-i k z} B^{(0) \dagger}_{ij}(y,z) \,.
\label{eq:Btwolegs}
\end{align}
This pattern continues to all orders with the $n$th vertex factor equal to $n$ Fourier transforms with momenta coordinates constrained to sum to $P$. Finally the generalization to spin is achieved by applying field indices to both sides of Eqs.~(\ref{eq:Aoneleg})-(\ref{eq:Btwolegs}).

\begin{figure*}[t]
\begin{center}
\label{fig:FVcorr}
\includegraphics[scale=0.45]{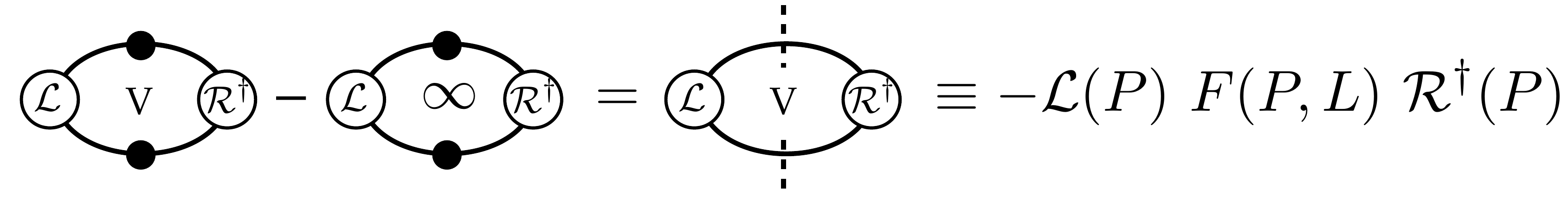}
\caption{ As discussed in the text, the difference between the finite and infinite volume two-particle loops is due to the cut of the loops, and can be written as the finite-volume matrix $F(P,L)$, defined in Eq.~(\ref{eq:FJJdef}), contracted with row and column vectors as shown.  }\label{fig:F_function}
\end{center}
\end{figure*}

With our general Feynman rules in hand, we next envision enumerating the full set of diagrams contributing to $C_L(P)$. We then organize this series into a simplified skeleton expansion. To do so one must identify all parts of diagrams with two-particles that can simultaneously go on-shell. Such particles must carry the total energy and momentum $E, \textbf P$ and must also have the quantum numbers of the operator $\mathcal B^\dagger$. Suppose that a given diagram has $N$ distinct pairs of propagators which carry $E, \textbf P$. Then this diagram is contained within the $N$th term of the following skeleton expansion [see also Figure~\ref{fig:F_function}]\footnote{A technical assumption first used here is that correlators and amplitudes, in both finite- and infinite-volume, are given by summing perturbation theory to all orders. \label{foot:scatamp}}
\begin{equation}
\begin{split}
\label{eq:CLskel}
C_L(P) & = \frac{1}{L^3}\sum_{\textbf k} \int \frac{d k_4}{2 \pi} A^{(0)}_{L,a}(k) \mathcal S_{a}(k) B^{(0) \dagger}_{L,a}(k) \\
& \hspace{30pt}+ \frac{1}{L^3}\sum_{\textbf k} \int \frac{d k_4}{2 \pi} \frac{1}{L^3}\sum_{\textbf k'} \int \frac{d k'_4}{2 \pi} \mathcal  \mathcal A^{(0)}_{L,a'}(k')  \mathcal S_{a'}(k') \mathcal M_{L,a',a}(k',k) \mathcal S_{a}(k) B^{(0)\dagger}_{L,a}(k) \,,
\end{split}
\end{equation}
where 
\begin{equation}
\label{eq:MLskel}
\mathcal M_{L,a',a}(k',k)   = -K_{a',a}(k',k) - \frac{1}{L^3}\sum_{\textbf p} \int \frac{d p_4}{2 \pi} K_{a',b}(k',p)  \mathcal S_b(p) \mathcal M_{L,b,a}(p,k) \,.
\end{equation}
Here
\begin{equation}
\mathcal S_{a}(k)   \equiv \xi_a \Delta^{s^a_1}_{a1}(k) \Delta^{s^a_2}_{a2}(P-k) \,,
\end{equation}
is the product of the two-propagators in the $a$th channel. For particles with nonzero spin, this should be understood as a direct product of matrices that reside in two distinct spaces. The propagators here are fully dressed and are defined in terms of the two-point function as
\begin{equation}
\label{eq:propdef}
\Delta^{s_1}_{a1}(k) \equiv \int \! d^4 x  e^{-i k  x} \langle 0 \vert T X^{s_1}_{a1}(x) X^{s_1,\mathrm{con}}_{a1}(0) \vert 0 \rangle \,,
\end{equation}
where $X^{s}$ is an interpolating field for a spin $s$ particle and $X^{s,\mathrm{con}}$ is the appropriate conjugate field. For example, for complex scalar fields $X^{0,\mathrm{con}} = X^{0 \dagger}$ and for Dirac fermions $X^{1/2,\mathrm{con}} = \overline{X^{1/2}}$. 

The functions $K_{a',a}(k',k)$ are two-to-two Bethe-Salpeter kernels with incoming particles in channel $a$ with momenta $k$, $P-k$, and outgoing particles in $a'$ with momenta $k', P-k'$. The Bethe-Salpeter kernel equals the sum of all diagrams that are two-particle irreducible with respect to propagator pairs carrying $(E, \textbf P)$. Equivalently, the kernels contain all diagrams with no on-shell intermediate states. Following Ref.~\cite{Kim:2005gf}, on can prove that for loop momenta in diagrams with no on-shell states, the difference between sums and integrals is exponentially suppressed as $e^{-mL}$. Since we neglect these corrections, we work throughout with the infinite-volume (integrated-loop-momenta) forms of the Bethe-Salpeter kernel $K_{a',a}$. We also use infinite-volume propagators in $\mathcal S_{a}$, which is justified since self-energy diagrams do not contain on-shell intermediate states. In short, all finite-volume effects arise from the sums over the two-particle loops shown explicitly in Eqs.~(\ref{eq:CLskel}) and (\ref{eq:MLskel}).

 Finally we explain the functions $A^{(0)}_{L,a'}(k')$ and $B^{(0)\dagger}_{L,a}(k)$, focusing our wording on the former. $A^{(0)}_{L,a'}(k')$ is the sum of all diagrams which contain exactly one insertion of a vertex factor, $\tilde A_{i\cdots}^{(0)}(k, \cdots)$, and two external legs. Additionally, the definition is restricted to only include diagrams which are two-particle irreducible in the $(E, \textbf P)$ channel of the outgoing two-particle pair. The final restriction demands further explanation and is best summarized through the condition
\begin{multline}
\label{eq:ALzerodef}
\int_L d^4 x \int_L d^4 y \ e^{- i P x} e^{i k y} \Big [ \langle 0 \vert \mathcal A(x) X^{s_1,\mathrm{con}}_{a1}(y) X^{s_2,\mathrm{con}}_{a2}(0) \vert 0 \rangle \Big ]_L [\mathcal S_{a}(k)]^{-1} =\\ A^{(0)}_{L,a}(k)+   \frac{1}{L^3}\sum_{\textbf k'} \int \frac{d k'_4}{2 \pi} \mathcal  \mathcal A^{(0)}_{L,a'}(k')  \mathcal S_{a'}(k') \mathcal M_{L,a',a}(k',k) \,.
\end{multline}
Note that the left-hand side simply defines the sum of every possible two-legged diagram with a single insertion of $\tilde A_{i\cdots}^{(0)}(k, \cdots)$. Thus, this relation defines $A^{(0)}_{L,a'}(k')$ as the partial sum of diagrams in which the two-particle scatterings, shown explicitly in the second term, are not included. 

Here we have allowed for the possibility that $A^{(0)}_{L,a'}(k')$ and $B^{(0)\dagger}_{L,a}(k)$ may have non-negligible finite-volume dependence. This will occur whenever some of diagrams within these quantities contain on-shell intermediate states. For the purposes of this work, we restrict attention to operators $\mathcal A$ and $\mathcal B^\dagger$ for which no such diagrams occur. We thus set $A^{(0)}_{L,a'}(k')$ and $B^{(0)\dagger}_{L,a}(k)$ equal to their infinite-volume counterparts and drop the $L$ subscript from now on. We revisit the restrictions on interpolating operators in Section \ref{sec:QCandME} below, where we make specific choices for the operators in order to derive our main results.  Since the diagrammatic representations of $A^{(0)}_{L,a'}(k')$ and $B^{(0)\dag}_{L,a}(k)$ will in general depend on the specific process in mind, in Figure~\ref{fig:examples_inB} we give examples of diagrams contributing to $B^{(0)\dag}_{L,a}(k)$ for a particular reaction, $N\gamma^*\rightarrow N\pi$.

We close this section my noting that spin has played a very minor role in the discussion. It is formally included everywhere above, and we stress in particular that that Eqs.~(\ref{eq:CLskel})-(\ref{eq:propdef}) include implicit field indices. The functions $A^{(0)}_{L,a'}$ and $B^{(0)\dagger}_{L,a}$ have one such index for each particle and the functions $K_{a',a}$, $\mathcal S_{a}$ have two for each. The indices are thus contracted in the natural way, with $A^{(0)}_{L,a'}$ viewed as a row vector, $K_{a',a}$ and $\mathcal S_a$ as matrices, and $B^{(0)\dagger}_{L,a}$ as a column. Similarly, the repeated channel indices are summed in all terms above. The new spin index structure and the modified forms of the propagators are the main differences that arise here as compared to the case of scalar particles. We will show how to accommodate this change in the next subsection.

%%%%%%%%%%%%%%%%%%%%%%%%%%%%%%%%%%%%%%%%%%%
%%%%%%%%%%%%%%%%%%%%%%%%%%%%%%%%%%%%%%%%%%%

\subsection{Finite-volume residue of two-particle loop}

Turning now to the second step listed above, in this subsection we identify the form of the finite-volume residue factor $F(P,L)$. The factor generically appears in expressions of the form 
\begin{align}
\label{eq:fvres}
\mathcal F_L & \equiv \sum_{a=1}^{N_c}  \left[\frac{1}{L^3}\sum_{\mathbf{k}}\hspace{-.5cm}\int~\right]\int \! \frac{d k_4}{2 \pi}  \mathcal L_{a}(P,k)  \mathcal S_{a}(k)  \mathcal R^\dagger_{a}(P,k) \,, \\ & = \sum_{a=1}^{N_c}  \bigg [ \frac{1}{L^3} \sum_{\textbf k} - \int \! \! \frac{d\textbf k}{(2 \pi)^3} \bigg ] \int \! \frac{d k_4}{2 \pi}  \mathcal L_{a}(P,k)  \mathcal S_{a}(k)  \mathcal R^\dagger_{a}(P,k) \,.
\end{align}
Here the functions $\mathcal L_a$ and $\mathcal R^\dagger_a$ can stand for either the interpolating functions $A^{(0)}$ and $B^{(0)\dagger}$ or the kernels $K$. This notation is only used within this subsection, and we rely on context to distinguish these functions from the residue matrix and the Lagrangian.  We now consider this finite-volume residue in three specific cases: two scalars, one scalar and one spin-half fermion, and two spin-half fermions. We then generalize the results for two arbitrary spin particles.

\subsubsection{Two scalar particles}
In the case of two scalar particles the finite-volume residue of the two-particle loop is given by
\begin{align}
\mathcal F_L & = \sum_{a=1}^{N_c} \xi_a \left[\frac{1}{L^3}\sum_{\mathbf{k}}\hspace{-.5cm}\int~\right] \int \frac{d k_4}{2 \pi} \ \mathcal L_a(P, k) \Delta^0_{a1}(k) \Delta^0_{a 2}(P-k)   \mathcal R^\dagger_a(P, k) \,, \\[5pt]
& = - \sum_{a=1}^{N_c} \xi_a  \left[\frac{1}{L^3}\sum_{\mathbf{k}}\hspace{-.5cm}\int~\right] \ \mathcal L_a(P, \textbf k^*_a)  \frac{1}{2 \omega_{a1} 2 \omega_{a 2}(E -   \omega_{a 1} - \omega_{a 2} + i \epsilon)}    \mathcal R^\dagger_a(P, \textbf k^*_a) \,.
\label{eq:exptwopartpole}
\end{align}
To reach the second equation here, we evaluated the integral over $k_4$ and discarded all contributions which are smooth functions of $\textbf k$. These give exponentially suppressed corrections, which we neglect throughout. We have also introduced new notation for the functions $\mathcal L_a$ and $\mathcal R_a$, indicating that these only depend on total momentum $P$ as well as the vector $\textbf k$ which we have boosted to the CM frame.

Next we use the identity proven in Ref.~\cite{Kim:2005gf} which states that one can make the replacement
\begin{equation}
\label{eq:onshellproj}
\mathcal L_a(P,\textbf k^*_a) \mathcal R^\dagger_a(P,\textbf k^*_a) \longrightarrow  \mathcal L_{a,l,m_l}(P) \left [4 \pi Y_{lm_l}(\hat {\textbf k}^*_a)  Y^*_{l'm_l'}(\hat {\textbf k}^*_a) \left(\frac{k^*_a}{q^*_a} \right)^{l+l'} \right ] \mathcal R^\dagger_{a',l',m'_l} (P) \,,
\end{equation}
where
\begin{align}
 \mathcal L_a(P, q^*_a \hat {\textbf k}^*_a)  & \equiv \mathcal L_{a,l,m_l}(P) \sqrt{4 \pi} Y_{lm_l}(\hat {\textbf k}^*_a) \,, \\
 \mathcal R^\dagger_{a'}(P, q^*_a \hat {\textbf k}^*_a)  & \equiv \mathcal R^\dagger_{a',l',m'_l}(P) \sqrt{4 \pi} Y^*_{l'm'_l}(\hat {\textbf k}^*_a) \,.
\end{align}
We deduce
\begin{equation}
\mathcal F_L = - \mathcal L_{a,l,m_l}(P)  F^{\mathrm{sc}}_{alm_l,a'l'm'_l} (P,L)\mathcal R^\dagger_{a',l',m'_l}(P)  \,,
\end{equation}
where $F^{\mathrm{sc}}_{alm_l,a'l'm'_l} (P,L)$ is defined in Eq.~(\ref{eq:Fscdef}) above. 

As mentioned above, for two-particles with momenta $\textbf k$ and $\textbf P-\textbf k$ subject to the constraint $E = \omega_{a1}+ \omega_{a2}$, the magnitude of a single particle's momentum in the CM frame must be $q^*_a$. Thus the key difference between the functions $ \mathcal L_a(P, k^*_a \hat {\textbf k}^*_a) $ and $ \mathcal L_a(P, q^*_a \hat {\textbf k}^*_a) $ is that the latter has been projected on-shell. This projection emerges because the sum-integral difference in Eq.~(\ref{eq:exptwopartpole}) is dominated by the two-particle pole. Equivalently, the difference between the left- and right-hand sides of Eq.~(\ref{eq:onshellproj}) regulates the pole in Eq.~(\ref{eq:exptwopartpole}). Thus the difference only generates exponentially suppressed corrections. Since we neglect these, the substitution is justified.

\subsubsection{One scalar and one spin-half fermion}

In the case of one scalar and one spin-half fermion the finite-volume residue takes the form
\begin{equation}
\label{eq:Fscalarfermion}
\mathcal F_L = - \sum_{a=1}^{N_c}  \left[\frac{1}{L^3}\sum_{\mathbf{k}}\hspace{-.5cm}\int~\right]  \mathcal L_{a}(P, \textbf k^*_a)   \left [ \frac{ \big  [\! - i \slashed k + m_{a1} \big ]_{k_4 = i \omega_{a1} }}{2 \omega_{a 1} 2 \omega_{a2}(E -  \omega_{a1} - \omega_{a2} + i \epsilon)}  \right ] \mathcal R^\dagger_{a}(P, \textbf k^*_a)  \,.
\end{equation}
This expression assumes that the propagating fermion has the quantum numbers of $\Psi \vert 0 \rangle$ (rather than $\overline \Psi \vert 0 \rangle$). The distinction enters through the sign of $k_4= i \omega_{a1}$ appearing in $\slashed k$. 

We now substitute the identity
\begin{equation}
\label{eq:spinordecom}
\big  [\!- i \slashed k + m_{a1} \big ]_{k_4 = i \omega_{a1} } = \sum_{\lambda_{1/2} = \pm} u_{a,\lambda_{1/2}}(\textbf k) \bar u_{a,\lambda_{1/2}}(\textbf k)  \,,
\end{equation}
where $u_{a,\lambda_{1/2}}(\textbf k)$ and $\bar u_{a,\lambda_{1/2}}(\textbf k)$ are spinors used to interpolate states with definite spin. Here we are using helicity states, but this identity holds for spinors $u$ and $\overline u$, which interpolate spin states quantized along any fixed direction in the CM frame of the single spin-half particle. We deduce
\begin{equation}
\label{eq:FLscalarfermion}
\mathcal F_L =  - \sum_{a=1}^{N_c} \xi_a \left[\frac{1}{L^3}\sum_{\mathbf{k}}\hspace{-.5cm}\int~\right] \mathcal L_{a,\lambda_{1/2}}(P, \textbf k^*_a)  \left [ \frac{ \delta_{\lambda_{1/2}\lambda_{1/2}'}  }{2 \omega_{a1} 2 \omega_{a2}(E -  \omega_{a1} - \omega_{a2} + i \epsilon )}  \right ] \mathcal R^\dagger_{a, \lambda_{1/2}'  }(P, \textbf k^*_a) \,,
\end{equation}
where
\begin{equation}
\label{eq:contractspinor}
\mathcal L_{a, \lambda_{1/2}}(P, \textbf k^*_a) \equiv \mathcal L_{a}(P, \textbf k^*_a)\ u_{a,\lambda_{1/2}}\!(\textbf k) \,, \ \ \ \ \ \ \mathcal R^\dagger_{a, \lambda_{1/2}}(P, \textbf k^*_a) \equiv  \bar u_{a, \lambda_{1/2}}\!(\textbf k)\  \mathcal R^\dagger_{a}(P, \textbf k^*_a) \,.
\end{equation}

Note that $\mathcal L_{a, \lambda_{1/2} }(P, \textbf k^*_a)$ and $\mathcal R^\dagger_{a, \lambda_{1/2} }(P, \textbf k^*_a)$ are the definite spin projections of $ \mathcal L_{a}(P, \textbf k^*_a)$ and $\mathcal R^\dagger_{a}(P, \textbf k^*_a)$. For example consider the special case 
\begin{equation}
 \mathcal L_{a}(P, \textbf k^*_a) = \lim_{k_4 \rightarrow i \omega_k, P_4 \rightarrow i E} \int d^4x d^4 y \,e^{i k x + i (P-k) y} \langle 0 \vert T \mathcal A(0) \Psi(x) \phi(y) \vert 0 \rangle \big [i \slashed k + m_{a1} \big ] \big [(P-k)^2 + m_{2a}^2 \big] \,.
\end{equation}
Then it follows that
\begin{equation}
\mathcal L_{a,  \lambda_{1/2} }(P, q^*_a \hat {\textbf k}^*_a)  = \langle 0 \vert \mathcal A(0) \vert E, \textbf P, \hat {\textbf k}^*_a, s^a, \lambda_{1/2} ,\ \mathrm{in} \rangle \,.
\end{equation}
The contracted spinor in Eq.~(\ref{eq:contractspinor}) is precisely the required element to convert the correlator (evaluated in the LSZ limit) to a matrix element involving a physical two-particle state with definite helicity. We next decompose in spherical harmonics to reach
\begin{align}
 \mathcal L_{a,\lambda_{1/2}}(P, q^*_a \hat {\textbf k}^*_a)  & = \mathcal L_{a,l,m_l,\lambda_{1/2}}(P) \sqrt{4 \pi} Y_{lm_l}(\hat {\textbf k}^*_a) \,, \\
 \mathcal R^\dagger_{a',\lambda_{1/2}'}(P, q^*_a \hat {\textbf k}^*_a)  & = \mathcal R^\dagger_{a',l',m'_l,\lambda_{1/2}'}(P) \sqrt{4 \pi} Y^*_{l'm'_l}(\hat {\textbf k}^*_a) \,.
\end{align}

As already discussed for general particles in Eqs.~(\ref{eq:ltoJtrans}) and (\ref{eq:CG}) above, these may also be defined in a basis of definite total angular momentum
\begin{gather}
\mathcal L_{a,J, M, l}(P) \equiv \sum_{m_l \lambda_{1/2}} \langle l \,m_l, \nicefrac 12 \, \lambda_{1/2} \vert J \, M \rangle \mathcal L_{a, l, m_l,\lambda_{1/2} }(P) \,, \\
\sum_{JM} \langle l \,m, \nicefrac 12 \, \lambda_{1/2} \vert J \, M \rangle  \mathcal L_{a,J, M, l }(P) \equiv \mathcal L_{a, l, m_l,\lambda_{1/2}}(P) \,.
\end{gather}
Applying the identity of Eq.~(\ref{eq:onshellproj}) used for scalar particles above, we deduce
\begin{align}
\mathcal F_L & =   - \mathcal L_{a, l, m_l,\lambda_{1/2}}(P)\  F_{a l m_l \lambda_{1/2},a' l' m_l' \lambda_{1/2}'}\!(P,L)\ \mathcal R^{\dagger}_{a', l', m_l',\lambda_{1/2}'}(P) \,, \\
 & =   - \mathcal L_{a,J, M, l}(P)\  F_{aJMl,a'J'M'l'}(P,L)\ \mathcal R^{\dagger}_{a',J', M', l'}(P) \,, 
\end{align}
where
\begin{align}
F_{a l m_l \lambda_{1/2},a' l' m_l' \lambda_{1/2}'}\!(P,L)  & \equiv  \delta_{\lambda_{1/2} \lambda_{1/2}'}  F^{\mathrm{sc}}_{alm_l,a'l'm_l'} (P,L) \,, \\
F_{aJMl,a'J'M'l'}(P,L) &  \equiv \sum_{m_l,\lambda_{1/2},m_l'} \langle l \,m, \nicefrac 12 \, \lambda_{1/2} \vert J M \rangle  \langle l' \,m_l', \nicefrac 12 \, \lambda_{1/2} \vert J' M' \rangle  F^{\mathrm{sc}}_{alm_l,a'l'm_l'} (P,L)   \,.
\end{align}
These are just the definitions of Eqs.~(\ref{eq:Flldef})-(\ref{eq:Fscdef}) above, applied to the special case of one scalar and one spin-half fermion.

We stress here that the only new ingredient for spin-half particles relative to scalars is the on-shell Dirac matrix, which can be trivially accommodated using Eq.~(\ref{eq:spinordecom}). The following points are crucial:
\begin{itemize}
\item It is justified to use the free form of the propagator in Eq.~(\ref{eq:Fscalarfermion}) only because of the sum-integral difference appearing in this equation. This allows one to replace fully dressed propagators with free propagators, because the difference between the two is a smooth function which thus gives only exponentially suppressed finite-volume corrections.
\item The observation that the on-shell Dirac matrix is equal to a sum of spinors, Eq.~(\ref{eq:spinordecom}), is equivalent to noting that it is simply the {\em identity} matrix when written in spin space. This point is a clear physical necessity: The propagation of free-particle states must be independent of the azimuthal spin of those states. 

\item  Using Eq.~(\ref{eq:spinordecom}) also makes it clear that the specific value of the matrix $\slashed k$ in a particular basis of the gamma matrices is irrelevant. For example, Refs.~\cite{Li:2012bi, Li:2014wga} first use symmetry arguments to set $\textbf k \cdot \vec \gamma$ to zero in Eq.~(\ref{eq:Fscalarfermion}). This simplification requires that the remaining factors in the summand are invariant under $\textbf k \rightarrow - \textbf k$, and thus only applies in the case of $\textbf P = 0$.
\end{itemize}

\subsubsection{Two spin one-half fermions}

In the case of two spin-half fermions the finite-volume residue takes the form
\begin{equation}
\mathcal F_L = - \sum_{a=1}^{N_c}  \left[\frac{1}{L^3}\sum_{\mathbf{k}}\hspace{-.5cm}\int~\right]  \mathcal L_{a,\alpha,\beta}(P, \textbf k^*_a)   \left [ \frac{ \big  [\! - i \slashed k + m_{a1} \big ]^{k_4 = i \omega_{a1}}_{\alpha \alpha'} \big  [\! - i (\slashed P - \slashed k) + m_{a2} \big ]^{P_4-k_4 = i \omega_{a2}}_{\beta \beta'}}{2 \omega_{a 1} 2 \omega_{a2}(E -  \omega_{a1} - \omega_{a2} + i \epsilon)}  \right ] \mathcal R^\dagger_{a,\alpha',\beta'}(P, \textbf k^*_a)  \,.
\end{equation}
Here we have explicitly shown the Dirac indices: $\alpha$, $\alpha'$, $\beta$, $\beta'$. We do so to stress that the two matrices in the numerator act on different spaces. Substituting the decomposition in spinors, Eq.~(\ref{eq:spinordecom}) above, we find
\begin{equation}
\mathcal F_L = - \sum_{a=1}^{N_c}  \left[\frac{1}{L^3}\sum_{\mathbf{k}}\hspace{-.5cm}\int~\right]  \mathcal L_{a,\lambda_{s_1},\lambda_{s_2}}(P, \textbf k^*_a)   \left [ \frac{\delta_{\lambda_{s_1} \lambda_{s_1} '}  \delta_{\lambda_{s_2} \lambda_{s_2} '}}{2 \omega_{a 1} 2 \omega_{a2}(E -  \omega_{a1} - \omega_{a2} + i \epsilon)}  \right ] \mathcal R^\dagger_{a,\lambda_{s_1}',\lambda_{s_2}'}(P, \textbf k^*_a)  \,,
\end{equation}
where
\begin{align}
  \mathcal L_{a,\lambda_{s_1},\lambda_{s_2}}(P, \textbf k^*_a)& \equiv\mathcal L_{a,\alpha,\beta}(P, \textbf k^*_a) \, u_{a,\alpha,\lambda_{s_1}}\!(\textbf k)\,   u_{a,\beta,\lambda_{s_2}}(\textbf P - \textbf k)  \,, \\
 \mathcal R^\dagger_{a,\lambda_{s_1}',\lambda_{s_2}'}(P, \textbf k^*_a) & \equiv \bar u_{a, \alpha',\lambda_{s_1}}\!(\textbf k) \, \bar u_{a, \beta',\lambda_{s_1}}(\textbf P -\textbf k)  \, \mathcal R^\dagger_{a,\alpha',\beta'}(P, \textbf k^*_a) \,.
\end{align}
In summary the helicity projection used above now appears twice for the case of two spin-half fermions. 

Decomposing in harmonics and applying the on-shell identity we find
\begin{align}
\mathcal F_L & =   - \mathcal L_{a, l, m_l,\lambda_{s_1},\lambda_{s_2}}\  F_{a l m_l \lambda_{s_1} \lambda_{s_2},a' l' m_l' \lambda_{s_1}' \lambda_{s_2}'}\!(P,L)\ \mathcal R^{\dagger}_{a', l', m_l',\lambda_{s_1}',\lambda_{s_2}'}(P) \,, \\
 & =   - \mathcal L_{a,J, M, l}\  F_{aJMl,a'J'M'l'}(P,L)\ \mathcal R^{\dagger}_{a',J', M', l'}(P) \,, 
\end{align}
where
\begin{align}
F_{a l m_l \lambda_{s_1} \lambda_{s_2},a' l' m_l' \lambda_{s_1}' \lambda_{s_2}'}\!(P,L)  & \equiv  \delta_{\lambda_{s_1} \lambda_{s_1}'}   \delta_{\lambda_{s_2} \lambda_{s_2}'}   F^{\mathrm{sc}}_{alm_l,a'l'm_l'} (P,L) \,, \\
F_{aJMlS,a'J'M'l'S'}(P,L)  & \equiv  \delta_{SS'}  \sum_{m_l,m'_l,m_S} \langle l \,m_l, S \, m_S \vert J M \rangle  \langle l' \,m_l', S' \, m_S \vert J' M' \rangle  F^{\mathrm{sc}}_{alm,a'l'm'}(P,L)    \,.
\end{align}
Again these are the definitions of Eqs.~(\ref{eq:Flldef})-(\ref{eq:Fscdef}) above, now applied to the special case of two spin-half fermions.

\subsubsection{General Spin}

Turning to general spin particles, we now make two key observations, true for all particle types, which allow us to reduce Eq.~(\ref{eq:fvres}). First we note that evaluating the $k_4$ integral and discarding terms which are exponentially suppressed generically gives an expression of the form
\begin{equation}
\mathcal F_L \equiv - \sum_{a=1}^{N_c} \xi_a \left[\frac{1}{L^3}\sum_{\mathbf{k}}\hspace{-.5cm}\int~\right]   \mathcal L_{a}(P,k) \frac{T_{a}(\textbf k)}{2 \omega_{a1} 2 \omega_{a 2}(E -   \omega_{a 1} - \omega_{a 2} + i \epsilon)} \mathcal R^\dagger_{a}(P,k) \,,
\end{equation}
where $T$ is a matrix with index space of the two-propagators, i.e.~two sets of field indices. Next we observe that
\begin{equation}
\mathcal L_{a}(P,k)  T_{a}(\textbf k) \mathcal R^\dagger_{a}(P,k) =\mathcal L_{a, \lambda_{s_1}, \lambda_{s_2}}(P,k) \ \delta_{\lambda_{s_1} \lambda_{s_1}'} \delta_{\lambda_{s_2} \lambda_{s_2}'} \ \mathcal R^\dagger_{a, \lambda_{s_1}',  \lambda_{s_2}'}(P,k) \,.
\end{equation}
Here the factors of $\mathcal R$ and $\mathcal L^\dagger$ appearing on the right-hand side are redefinitions of the original $\mathcal L$ and $\mathcal R^\dagger$ appearing on the left which couple to states with definite $\lambda_{s_1}$, $\lambda_{s_2}$. Further, these objects are normalized so that the infinite summations in the next section will give standard infinite-volume matrix elements and scattering amplitudes. As already mentioned above, we know that $T$ must be proportional to the identity matrix for particles of any spin. This follows from the physical observation that for free-particle states the propagator cannot vary for different azimuthal components.

Combining this result with the on-shell projection described by Eq.~(\ref{eq:onshellproj}), we deduce
\begin{equation}
\mathcal F_L \equiv - \mathcal L(P) F(P,L) \mathcal R^\dagger(P) \,,
\end{equation}
where $\mathcal L(P)$ is understood as a row-vector and $\mathcal R^\dagger(P)$ as a column vector with indices of either $\{l\}$ or $\{J\}$. $F(P,L)$ is defined in Eqs.~(\ref{eq:Flldef})-(\ref{eq:Fscdef}) and must be used here with the same basis as $\mathcal L(P)$ and $\mathcal R^\dagger(P)$.

%%%%%%%%%%%%%%%%%%%%%%%%%%%%%%%%%%%%%%%%%%%

\begin{figure*}[t]
\begin{center}
\label{fig:FVcorr}
\includegraphics[scale=0.45]{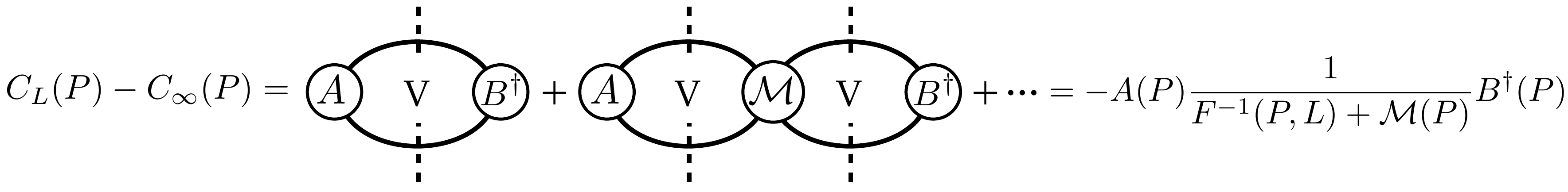}
\caption{ As discussed in Section~\ref{sec:sumdiags}, the difference between the finite and infinite volume correlation function is a geometric series in $-\mathcal M(P) F(P,L)$, where $\mathcal M(P)$ is the scattering amplitude and $F(P,L)$ is the finite volume function defined in Eq.~(\ref{eq:FJJdef}). }\label{fig:CV_m_Cinfy}
\end{center}
\end{figure*}

%%%%%%%%%%%%%%%%%%%%%%%%%%%%%%%%%%%%%%%%%%%
\subsection{Summation of diagrams~\label{sec:sumdiags}}

We now apply the key identity of the previous section
\begin{equation}
 \mathcal F_L \equiv \sum_{a=1}^{N_c}  \left[\frac{1}{L^3}\sum_{\mathbf{k}}\hspace{-.5cm}\int~\right]\int \! \frac{d k_4}{2 \pi}  \mathcal L_{a}(P,k)  \mathcal S_{a}(k)  \mathcal R^\dagger_{a}(P,k) = - \mathcal L(P) F(P,L) \mathcal R^\dagger(P) \,,
\end{equation}
to the skeleton expansion given in Eqs.~(\ref{eq:CLskel}) and (\ref{eq:MLskel}). Each sum is replaced with an integral plus a residue which contains $F$. Reorganizing by number of $F$ insertions, as shown in Figure~\ref{fig:CV_m_Cinfy}, we deduce
\begin{equation}
C_L(P) = C_\infty(P) + \sum_{n=0}^\infty A(P) [-F(P,L)] [-\mathcal M(P) F(P,L)]^n B^{\dagger}(P) \,,
\end{equation}
and thus conclude Eq.~(\ref{eq:CLres}).  

As already mentioned in Footnote \ref{foot:scatamp}, this result assumes that the correct scattering amplitude and matrix elements are given by summing the perturbative definitions to all orders. In particular the quantities $A_{\{l\}}(P)$, $B_{\{l'\}}^{\dagger}(P)$ and $\mathcal M_{\{l\}, \{l'\}}(P)$ are reached by first defining the functions
\begin{align}
A(P,k) &\equiv A^{(0)}(P,k)+\int \! \! \frac{d^4 k'}{(2 \pi)^4} A^{(0)}(P,k')  \mathcal S (k') \mathcal M(P,k',k)\,, \\
B^{\dagger}(P,k') & \equiv B^{(0)\dagger}(P,k')+\int \! \! \frac{d^4 k}{(2 \pi)^4}  \mathcal M(P,k',k)  \mathcal S (k) B^{(0)\dagger}(P,k)\,, \\
\mathcal M(P,k',k)  & \equiv -K(k',k) -  \int \! \! \frac{d^4 p}{(2 \pi)^4} K(k',p)  \mathcal S (p) \mathcal M(P,p,k) \,,
\end{align}
which each have implicit indices $a,s^a_{1},\lambda_{s_1},s^a_{2},\lambda_{s_2}$. Projecting the vectors $k$ and $k'$ on-shell reduces the coordinate dependence to the unit vectors $\hat {\textbf k}^*_a$ and $\hat {\textbf k}'^*_a$. Finally, decomposing these in spherical harmonics gives $A_{\{l\}}(P)$, $B_{\{l'\}}^{\dagger}(P)$ and $\mathcal M_{\{l\}, \{l'\}}(P)$.

\section{Quantization Condition and Matrix element relations}

\label{sec:QCandME}

In this section we show how Eq.~(\ref{eq:CLres}) can be used to derive a relation between finite-volume spectrum and scattering amplitudes, as well as relations between finite- and infinite-volume matrix elements. The main results of this section, and of the paper, are the matrix element relations, Eqs.~(\ref{eq:onetotwomain}), (\ref{eq:onetotwosign}), (\ref{eq:zerototwomain}) and (\ref{eq:zerototwosign}).

Beginning with the relation between spectrum and scattering, we note that poles in $C_L(P)$ are located at the energies of the finite-volume theory. More precisely, poles are located at $P_4 = i E_{n,\textbf P, L}$ where $E_{1,\textbf P, L}, E_{2,\textbf P, L}, \cdots$ is the finite-volume spectrum at fixed $\textbf P, L$. Since no poles appear in $C_\infty(P)$, $A(P)$, or $B^{\dagger}(P)$ the spectrum is given by all energies for which the matrix
\begin{equation}
\frac{1}{F^{-1}(P,L) + \mathcal M(P)} \,,
\end{equation}
has a divergent eigenvalue. Equivalently the spectrum is given by all zeroes of the function
\begin{equation}
\label{eq:QCinSecFour}
\Delta(P,L) \equiv \det[F^{-1}(P,L) + \mathcal M(P)] \,.
\end{equation}
This is the fully general two-particle quantization condition which was first presented in Ref.~\cite{Briceno:2012yi}.

We now turn to relating finite- and infinite-volume matrix elements. To accomplish this we consider the Fourier transformed correlator
\begin{align}
C_L(x_4-y_4, \textbf P) & \equiv \int_L \! d \textbf x \int_L \! d \textbf y \ e^{- i \textbf P \cdot (\textbf x - \textbf y)} \Big [ \langle 0 \vert T \mathcal A(x) \mathcal B^\dagger(y) \vert 0 \rangle \Big ]_L \,, \\
& = \int_L \! d \textbf x \int_L \! d \textbf y \ e^{- i \textbf P \cdot (\textbf x - \textbf y)} \sum_n \Big [ \langle 0 \vert \mathcal A(x_4, \textbf x)  \vert E_n, \textbf P, L \rangle \Big ]_L \Big [ \langle E_n, \textbf P, L \vert \mathcal B^\dagger(y_4, \textbf y) \vert 0 \rangle \Big ]_L\,, \\
& =  \int_L \! d \textbf x \int_L \! d \textbf y \ \sum_n e^{- E_{n}(x_4-y_4)}  \Big [  \langle 0 \vert  \mathcal A(0)  \vert E_n, \textbf P, L \rangle \Big ]_L \Big [ \langle E_n, \textbf P, L \vert \mathcal B^\dagger(0) \vert 0 \rangle \Big ]_L \,, \\
& =  L^6 \sum_n e^{- E_{n}(x_4-y_4)}   \Big [ \langle 0 \vert  \mathcal A(0)  \vert E_n, \textbf P, L \rangle \Big ]_L \Big [ \langle E_n, \textbf P, L \vert \mathcal B^\dagger(0) \vert 0 \rangle \Big ]_L \,.
\label{eq:completeset}
\end{align}
In the second step we have assumed $x_4>y_4$ and have inserted a complete set of normalized finite-volume states. As a result of the Fourier transform, it is sufficient to insert only states with total momentum $\textbf P$. In the third step we have pulled out the time-evolution and translation operators. The time evolution operators give the exponential time dependence and the translation operators cancel the phase factors from the Fourier transforms. We thus conclude that the integrand does not depend on $\textbf x$ or $\textbf y$, so that the integrals simply give factors of volume as shown in the final step.

We now evaluate the same correlator in a different approach, by using Eq.~(\ref{eq:CLres})
\begin{align}
C_L(x_4-y_4, \textbf P)& \equiv L^3 \int \frac{d P_4}{2 \pi} e^{i P_4(x_4-y_4)} C_L(P) \,, \\
&= L^3 \int \frac{d P_4}{2 \pi} e^{i P_4(x_4-y_4)} \left[ C_{\infty}(P) - A(P) \frac{1}{F^{-1}(P,L) + \mathcal M(P)} B^{\dagger}(P) \right ] \,, \label{eq:steptwo} \\
& = \sum_n e^{- E_{n}(x_4-y_4)}L^3  \langle 0 \vert   {\mathcal A}(0) \vert E_{n}, \textbf P, \{J\}, \mathrm{in} \rangle \Big [ \mathcal R_{\{J\}, \{J'\}}(E_{n}, \textbf P)\Big ] \langle E_{n}, \textbf P,\{J'\},\mathrm{out} \vert \mathcal B^\dagger(0) \vert 0\rangle
\label{eq:stepthree} \,. 
\end{align}
where
\begin{equation}
\mathcal R_{\{J\},\{J'\}}(E_{n}, \textbf P) \equiv  \lim_{P_4 \rightarrow i E_{n}} \left[ - (i P_4 + E_{n}) \frac{1}{F^{-1}(P,L) + \mathcal M(P)}\right]_{\{J\},\{J'\}} \,,
\label{eq:Resfunc}\end{equation}
is the residue of the matrix that appears between $A(P)$ and $B^{\dagger}(P)$, evaluated at the $n$th two-particle energy. To go from Eq.~(\ref{eq:steptwo}) to Eq.~(\ref{eq:stepthree}) we assumed that $x_4>y_4$, allowing to close the contour in the upper half of the complex $P_4$ plane. The only analytic structure encircled in the contour is the tower of finite-volume-spectrum poles along the positive imaginary axis. The integral thus reduces to a sum of residues at these poles.

 %%%%%%%%%%%%%%%%%%%%%%%%%%%%%%%%%%%%%%%%%%%
\begin{figure*}[t]
\begin{center}
\subfigure[]{
\label{fig:inB}
\includegraphics[scale=0.5]{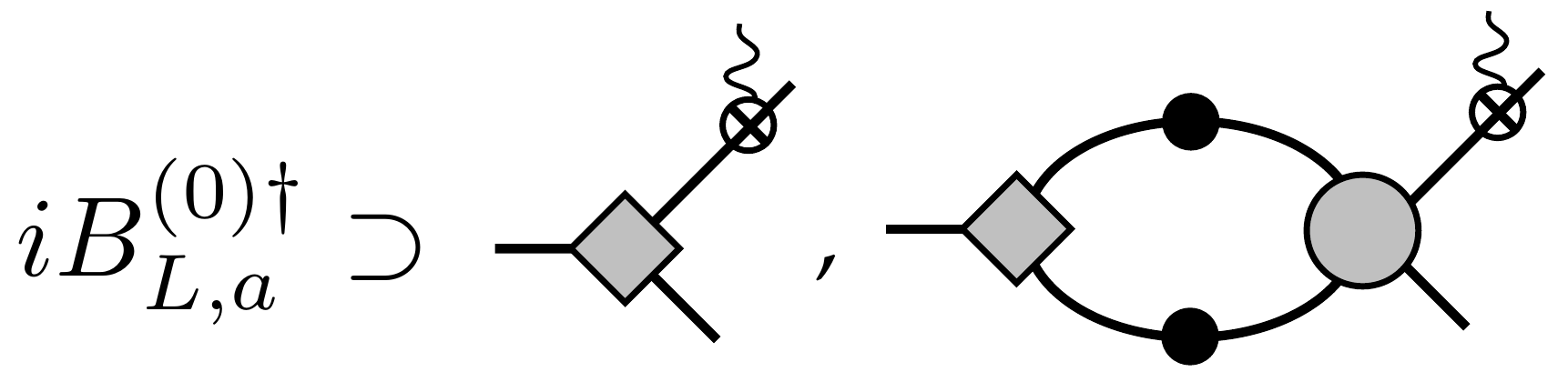}}
~~~~~~\subfigure[]{
\label{fig:notinB}
\includegraphics[scale=0.5]{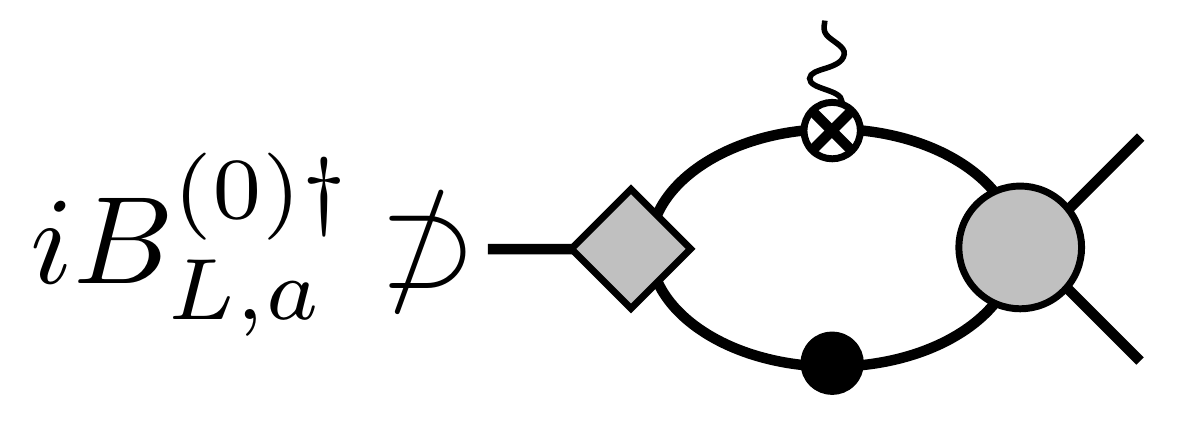}} 
\caption{
Shown are examples of diagrams that are present (a) as well as those that do not contribute (b) to $iB^{(0)\dag}_{L,a}$ for, for example, $N\gamma^*\rightarrow N\pi$. Equation~(\ref{eq:ALzerodef}) defines $iB^{(0)\dag}_{L,a}$ for generic systems.
}\label{fig:examples_inB}
\end{center}
\end{figure*}
%%%%%%%%%%%%%%%%%%%%%%%%%%%%%%%%%%%%%%%%%%%

Equating Eqs.~(\ref{eq:completeset}) and (\ref{eq:stepthree}) we deduce
\begin{multline}
\label{eq:firstmaster}
\Big [ \langle 0 \vert    {\mathcal A}(0) \vert E_n, \textbf P, L \rangle \Big ]_L \Big [ \langle E_n, \textbf P, L  \vert  {\mathcal B}^\dagger(0) \vert 0\rangle \Big ]_L=\\
  \frac{1}{L^3}   \langle 0 \vert    {\mathcal A}(0) \vert E_{n }, \textbf P, \{J\}, \mathrm{in} \rangle \Big [ \mathcal R_{\{J\}, \{J'\}}(E_{n}, \textbf P)\Big ] \langle E_{n }, \textbf P,\{J'\},\mathrm{out} \vert  {\mathcal B}^\dagger(0) \vert 0\rangle  \,.
\end{multline}

From this result follows all matrix element relations presented in this work [Eqs.~(\ref{eq:onetotwomain}), (\ref{eq:onetotwosign}), (\ref{eq:zerototwomain}) and (\ref{eq:zerototwosign}) below]. The equality relates matrix elements between finite-volume states and two-particle asymptotic states. We stress here that the result is only valid for $E_n^* = [E_n^2 - \textbf P^2]^{1/2}$ below the lowest three or four-particle threshold. This restriction arrises because it is only for the energy poles below inelastic threshold that Eq.~(\ref{eq:CLres}) is valid. Thus, although we know that the correlator in Eqs.~(\ref{eq:completeset}) and (\ref{eq:stepthree}) must equal an infinite series of decaying exponentials, we only quantitatively control the exponentials corresponding to two-particle states. We stress however that matching between (\ref{eq:completeset}) and (\ref{eq:stepthree}) is unambiguous for each coefficient of the Euclidean-time-dependent exponentials. For this reason, as long as $E^*_n$ is below multi-particle threshold, then Eq.~(\ref{eq:firstmaster}) includes all power-law finite-volume effects and only ignores exponentially suppressed corrections of the form $e^{-mL}$.

Before deriving our main results from Eq.~(\ref{eq:firstmaster}), we consider an alternative form of the relation. We begin by taking the ratio of the equation with a slightly modified version in which ${\mathcal B}^\dagger$ is replaced with ${\mathcal A}^\dagger$. This gives
\begin{equation}
\label{eq:ratiostart}
\frac{ \Big [  \langle E_n, \textbf P, L \vert    {\mathcal A}^\dagger(0) \vert 0 \rangle \Big ]_L }{ \Big [ \langle E_n, \textbf P, L \vert    {\mathcal B}^\dagger(0) \vert 0 \rangle \Big ]_L } =
    \frac{ \langle 0 \vert    {\mathcal A}(0) \vert E_{n}, \textbf P, \{J\}, \mathrm{in} \rangle \Big [ \mathcal R_{\{J\}, \{J'\}}(E_{n}, \textbf P)\Big ] \langle E_{n}, \textbf P,\{J'\},\mathrm{out} \vert  {\mathcal A}^\dagger(0) \vert 0\rangle }{\langle 0 \vert    {\mathcal A}(0) \vert E_{n}, \textbf P, \{J\}, \mathrm{in} \rangle \Big [ \mathcal R_{\{J\}, \{J'\}}(E_{n}, \textbf P)\Big ] \langle E_{n}, \textbf P,\{J'\},\mathrm{out} \vert  {\mathcal B}^\dagger(0) \vert 0\rangle} \,.
\end{equation}
Next we recall a result demonstrated in Ref.~\cite{Briceno:2014uqa}, that the matrix $\mathcal R$ has only one-nonzero eigenvalue, and is thus equal to an outerproduct of vectors
\begin{equation}
\label{eq:outerproduct}
\mathcal R_{\{J\}, \{J'\}}(E_{n }, \textbf P) \equiv  \mathcal {E}^{\mathrm{in}}_{\{J\}} \mathcal {E}^{\mathrm{out}}_{\{J'\}}\,,
\end{equation}
where $\mathcal{E}^{\mathrm{in}}_{\{J\}}$ is understood as a column vector in space of open channels, and $\mathcal{E}^{\mathrm{out}}_{\{J\}}$ as a row vector. Substituting this into Eq.~(\ref{eq:ratiostart}) gives 
\begin{align}
\label{eq:ratiomaster}
\frac{\Big [ \langle E_n, \textbf P, L \vert    {\mathcal A}^\dagger(0) \vert 0 \rangle \Big ]_L }{\Big [ \langle E_n, \textbf P, L \vert    {\mathcal B}^\dagger(0) \vert 0 \rangle \Big ]_L}  & =
 \frac{\mathcal {E}^{\mathrm{out}}_{\{J'\}} \ \langle E_{n }, \textbf P,\{J'\},\mathrm{out} \vert  {\mathcal A}^\dagger(0) \vert 0\rangle }{ \mathcal {E}^{\mathrm{out}}_{\{J'\}} \ \langle E_{n }, \textbf P,\{J'\},\mathrm{out} \vert  {\mathcal B}^\dagger(0) \vert 0\rangle} \,, \\
 &=
    \frac{ \mathcal X^\dagger_{\{J\}} \Big [ \mathcal R_{\{J\}, \{J'\}}(E_{n }, \textbf P)\Big ] \langle E_{n }, \textbf P,\{J'\},\mathrm{out} \vert  {\mathcal A}^\dagger(0) \vert 0\rangle }{\mathcal X^\dagger_{\{J\}} \Big [ \mathcal R_{\{J\}, \{J'\}}(E_{n }, \textbf P)\Big ] \langle E_{n }, \textbf P,\{J'\},\mathrm{out} \vert  {\mathcal B}^\dagger(0) \vert 0\rangle} \,.
\end{align}
To reach the second equality we have multiplied the numerator and denominator by $ \mathcal X^\dagger_{\{J\}} \mathcal{E}^{\mathrm{in}}_{\{J\}}$ where $\mathcal X^\dagger_{\{J\}}$ is a completely arbitrary vector that can be freely chosen to make the relation as convenient as possible. The advantage of the second equality is that, although one can readily show that $\mathcal R$ can be written as an outerproduct, it can be complicated to determine the specific forms of $\mathcal{E}^{\mathrm{in}}_{\{J\}}$ and $\mathcal{E}^{\mathrm{out}}_{\{J'\}}$.

\subsection{One-to-two transition amplitudes with arbitrary spin}
In Ref.~\cite{Briceno:2014uqa}, we demonstrated how to relate matrix elements of external currents to transition amplitudes by formally determining two-point and three-point functions and then taking appropriate ratios. In that earlier work we also discussed in great detail the dependence of the correlation functions on off-shell scattering amplitudes. In this section we circumvent a great deal of the discussion of our previous work and use Eqs.~(\ref{eq:firstmaster}) and (\ref{eq:ratiomaster}) to generalize the matrix element relations to arbitrary spin. Beginning with (\ref{eq:firstmaster}), we note that one can formally choose the operators $\mathcal A$ and $\mathcal B^\dagger$ to satisfy
\begin{align}
\label{eq:Bchoiceone}
{\mathcal B}^\dagger(x) & = \frac{1}{\sqrt{2 E_{0, \textbf P} L^3}} \mathcal J_A(x) \lim_{P_4 \rightarrow i E_{0, \textbf P}} [P^2 + M^2]  \int d^4 y e^{i P y} \Phi^\dagger(x+y)  \,, \\
{\mathcal A}(x) & = \frac{1}{\sqrt{2 E_{0, \textbf P} L^3}} \lim_{P_4 \rightarrow i E_{0, \textbf P}} [P^2 + M^2]  \int d^4 y e^{-i P y} \Phi(x+y)  \mathcal J^\dagger_A(x) \,.
\label{eq:Achoiceone}
\end{align}
where  $\mathcal J_A$ is an arbitrary local current and $\Phi$ is the interpolating field for a single particle which is stable under the interactions governed by the Lagrangian, Eq.~(\ref{eq:lagdef}). 

Here it is important to note that these specific choices for the operators satisfy the general forms assumed to reach Eq.~(\ref{eq:CLres}). To see this we must consider the endcap functions $A^{(0)}_{L,a'}(k')$ and $B^{(0)\dagger}_{L,a}(k)$, defined in Eq.~(\ref{eq:ALzerodef}) above. For this discussion we actually define the endcap functions in terms of $\sqrt{2 E_{0, \textbf P} L^3} \mathcal A$ and $\sqrt{2 E_{0, \textbf P} L^3} \mathcal B^\dagger$ so that the prefactors in Eqs.~(\ref{eq:Bchoiceone}) and (\ref{eq:Achoiceone}) do not confuse the arguments. For the specific choice of Eq.~(\ref{eq:Bchoiceone}), $B^{(0)\dagger}_{L,a}(k)$ is equal to the sum of all one-to-two diagrams with the outgoing two-to-two finite volume scatterings amputated {[see Figure~\ref{fig:examples_inB} for examples and a counterexample]}. All contributing diagrams have no on-shell intermediate states and it thus follows that $B^{(0)\dagger}_{L,a}(k)$ is equal to its infinite-volume form, up to exponentially suppressed corrections. To see that this is the case, one must discount on-shell states both before and after the current insertion. Those appearing before are ruled out by the assumption that the incoming particle is stable. Those after the current must be two-particle states due to the restriction on total four-momentum. These are precisely the states that are amputated in the definition of $B^{(0)\dagger}_{L,a}(k)$. Finally, for $A^{(0)}_{L,a'}(k')$ two-to-one diagrams must be considered. Applying the same arguments, we deduce that no on-shell states appear and thus that the finite-volume corrections to both objects are negligible.

Returning to Eq.~(\ref{eq:Bchoiceone}) and (\ref{eq:Achoiceone}), the Fourier transform, propagator amputation and on-shell limit are the prescription required so that the operator picks out a one-particle state. In finite volume we find
\begin{equation}
\Big [ \langle E_n, \textbf P, L \vert    {\mathcal B}^\dagger(0) \vert 0 \rangle \Big ]_L =  \langle E'_n, \textbf P', L  \vert \mathcal J_A(0) \vert E_{0}, \textbf P,L ,1\rangle  \,,
\end{equation}
where the finite-volume one-particle state is normalized to unity
\begin{equation}
\langle E_0, \textbf P,L ,1\vert E_0, \textbf P,L ,1 \rangle = 1\,.
\end{equation}
In infinite volume the result is
\begin{equation}
\langle E_{n}, \textbf P,\{J'\},\mathrm{out} \vert  {\mathcal B}^\dagger(0) \vert 0\rangle = \frac{1}{\sqrt{2 E_{0, \textbf P} L^3}}    \langle E_{n}, \textbf P,\{J'\},\mathrm{out} \vert \mathcal J_A(0) \vert E_{0, \textbf P} , \textbf P, \{J\}_{1} \rangle
\end{equation}
where the infinite-volume states satisfy standard relativistic normalization
\begin{equation}
 \langle E_{0, \textbf P'} , \textbf P', \{J\}_{1} \vert E_{0, \textbf P} , \textbf P, \{J\}_{1} \rangle = 2 E_{0, \textbf P} (2 \pi)^3 \delta^3(\textbf P - \textbf P') \,, 
\end{equation}
see also Eq.~(\ref{eq:twopartnorm}) above.  

Substituting Eqs.~(\ref{eq:Bchoiceone}) and (\ref{eq:Achoiceone}) into Eq.~(\ref{eq:firstmaster}), we deduce 
\begin{equation} \vert \langle E'_n, \textbf P', L  \vert \mathcal J_A(0) \vert E_{0}, \textbf P,L ,1\rangle \vert  = \frac{1}{L^3} \frac{1}{\sqrt{2E_{0,\textbf P}} }  \sqrt{\mathcal H^{\mathrm{in}}_{A,\{J\}} \Big [ \mathcal R_{\{J\}, \{J\}'}(E_{n}', \textbf P')\Big ]  \mathcal H^{\mathrm{out}}_{A,\{J'\}} } \,,
\end{equation}
where
\begin{equation}
 \mathcal H^{\mathrm{out}}_{A,\{J'\}}(E_{0,\textbf P}, \textbf P; E_{n}', \textbf P')  \equiv  \langle E'_{n}, \textbf P',\{J'\},\mathrm{out} \vert \mathcal J_A(0)\vert E_{0,\textbf P} , \textbf P,\{J\}_1   \rangle \,.
\end{equation}
This can be trivially rewritten with all operators in momentum space
\begin{equation}
\label{eq:onetotwomain}
\vert \langle E'_n, \textbf P', L  \vert \widetilde {\mathcal J}_A(0, \textbf P - \textbf P') \vert E_{0}, \textbf P,L ,1\rangle \vert  = \frac{1}{\sqrt{2E_{0,\textbf P}}} \sqrt{\mathcal H^{\mathrm{in}}_{A,\{J\}} \Big [ \mathcal R_{\{J\}, \{J\}'}(E_{n}', \textbf P')\Big ]  \mathcal H^{\mathrm{out}}_{A,\{J'\}} } \,,
\end{equation}
where
\begin{equation}
\big[ \mathcal H^{\mathrm{out}}_{A,\{J'\}}(E_{0,\textbf P}, \textbf P; E_{n}', \textbf P') \big] \ (2 \pi)^3  \delta^3(\textbf P - \textbf P' - \textbf Q) \equiv  \langle E'_{n}, \textbf P',\{J'\},\mathrm{out} \vert \widetilde {\mathcal J}_A(0, \textbf Q)\vert E_{0,\textbf P} , \textbf P,\{J\}_1   \rangle \,,
\end{equation}
and the operator $\widetilde{\mathcal{J}}_A$ is defined as the Fourier transform of the position space current, ${\mathcal{J}}_A$,
\begin{eqnarray}
\widetilde{\mathcal{J}}_A(x_0,\textbf{Q})&=&\int d \textbf x~e^{-i\textbf{Q}\cdot \textbf{x}} \mathcal{J}_A(x).
\end{eqnarray}
This is the most general possible Lellouch-L\"uscher relation for cubic volumes and two-particle states. We give an example of the utility of this result in Section~\ref{sec:NpiSPwave}, and in Appendices \ref{sec:free_limit} and \ref{sec:narrow_width} we discuss the free and narrow-width limits, respectively. 

Note next that Eq.~(\ref{eq:onetotwomain}) does not allow one to access the signs of transition amplitudes. Although the absolute sign of a given matrix element is not physically observable, the relative sign between two matrix elements is. With this in mind, we take Eq.~(\ref{eq:ratiomaster}) and substitute
\begin{align}
{\mathcal A}^\dagger(0) & = \frac{1}{\sqrt{2 E_{0, \textbf P} L^3}} \mathcal J_{A_1}(0) \lim_{P_4 \rightarrow i E_{0, \textbf P}} [P^2 + M^2]  \int d^4 x e^{i P x} \Phi^\dagger(x) \,, \\
{\mathcal B}^\dagger(0) & = \frac{1}{\sqrt{2 E_{0, \textbf P} L^3}} \mathcal J_{A_2}(0) \lim_{P_4 \rightarrow i E_{0, \textbf P}} [P^2 + M^2]  \int d^4 x e^{i P x} \Phi^\dagger(x)\,.
\end{align}
This gives
\begin{align}
\label{eq:onetotwosign}
\frac{\langle E'_n, \textbf P', L  \vert \widetilde {\mathcal J}_{A_1}(0, \textbf P - \textbf P') \vert E_{0}, \textbf P,L ,1\rangle}{\langle E'_n, \textbf P', L  \vert \widetilde {\mathcal J}_{A_2}(0, \textbf P - \textbf P') \vert E_{0}, \textbf P,L ,1\rangle}  & = \frac{\mathcal{E}^{\mathrm{out}}_{\{J'\}}\  \mathcal H^{\mathrm{out}}_{A_1,\{J'\}} }{ \mathcal{E}^{\mathrm{out}}_{\{J'\}} \  \mathcal H^{\mathrm{out}}_{A_2,\{J'\}} }  \,, \\[5pt]
& = \frac{\mathcal X^\dagger_{\{J\}} \Big [ \mathcal R_{\{J\}, \{J'\}}(E_{n}', \textbf P')\Big ]  \mathcal H^{\mathrm{out}}_{A_1,\{J'\}} }{\mathcal X^\dagger_{\{J\}} \Big [ \mathcal R_{\{J\}, \{J'\}}(E_{n}', \textbf P')\Big ]  \mathcal H^{\mathrm{out}}_{A_2,\{J'\}} }\,,
\end{align}
where we used that in the ratio of finite-volume matrix elements, one can replace the position space current with a momentum space current, since these only differ by a factor of $L^3$. 

Eq.~(\ref{eq:onetotwosign}) constrains the relative sign and may be more useful for other reasons in the analysis of a given physical system. In general, $ \mathcal H ^{\mathrm{out}}_{A_j,\{J'\}}$ is an infinite vector over all open channels and all partial waves that mix in accordance with the symmetry of the system. If one ignores all but one entry, it is evident that there is a one-to-one mapping between the relative sign of finite-volume matrix elements and that between infinite-volume transition amplitudes. When including more than one partial wave or particle channel, Eqs.~(\ref{eq:onetotwomain}) and (\ref{eq:onetotwosign}) allow one to simultaneously constrain the absolute value and relative sign of these transition amplitudes. This procedure is analogous to that which was implemented in a recent LQCD study of $K\pi-K\eta$~\cite{Dudek:2014qha, Wilson:2014cna}.

\subsection{Vacuum to two-particle transition amplitudes with arbitrary spin~\label{sec:vac_to_two}}

An even more straightforward result of Eq.~(\ref{eq:firstmaster}) above is reached by setting
\begin{align}
{\mathcal B}^\dagger(0) & \equiv \mathcal J_A(0) ,\hspace{1cm}
 {\mathcal A} (0)   \equiv   \mathcal J^\dagger_A(0 )  \,,
\end{align}
where, once again, $\mathcal J_A$ is a local current. 
One finds
\begin{equation}
\label{eq:zerototwomain}
\vert \langle E_n, \textbf P, L \vert \widetilde {\mathcal J}_A(0, -\textbf P) \vert 0 \rangle \vert  = \sqrt{\mathcal V^{\mathrm{in}}_{A,\{J\}}  \Big[ L^3 \mathcal R_{\{J\}, \{J'\}}(E_{n}, \textbf P)\Big ]  \mathcal V^{\mathrm{out}}_{A,\{J'\}} } \,, 
\end{equation}
where
\begin{align}
 \mathcal V^{\mathrm{out}}_{A,\{J'\}}(E_{n}, \textbf P) & \equiv \langle E_{n}, \textbf P,\{J'\},\mathrm{out} \vert  {\mathcal J}_A(0)\vert 0  \rangle \,, \\
\big [ \mathcal V^{\mathrm{out}}_{A,\{J'\}}(E_{n}, \textbf P) \big ]\ (2 \pi)^3  \delta^3(\textbf P + \textbf Q) & \equiv \langle E_{n}, \textbf P,\{J'\},\mathrm{out} \vert \widetilde {\mathcal J}_A(0, \textbf Q)\vert 0  \rangle \,.
\end{align}

Similarly, substituting
\begin{align}
{\mathcal A}^\dagger(0)  & \equiv \mathcal J_{A_1}(0)   \,, 
\hspace{1cm}
{\mathcal B}^\dagger(0)   \equiv \mathcal J_{A_2}(0)  \,,
\end{align}
into Eq.~(\ref{eq:ratiomaster}) gives 
\begin{align}
\frac{\langle E_n, \textbf P, L  \vert \widetilde {\mathcal J}_{A_1}(0, - \textbf P) \vert 0 \rangle}{\langle E_n, \textbf P, L  \vert \widetilde {\mathcal J}_{A_2}(0,  - \textbf P) \vert 0 \rangle}  & = \frac{\mathcal E^{\mathrm{out}}_{\{J'\}} \  \mathcal V^{\mathrm{out}}_{A_1,\{J'\}} }{ \mathcal E^{\mathrm{out}}_{ \{J'\}} \  \mathcal V^{\mathrm{out}}_{A_2,\{J'\}} } \nn \,, \\[5pt]
\label{eq:zerototwosign}
& = \frac{\mathcal X^\dagger_{\{J\}} \Big [ \mathcal R_{\{J\}, \{J'\}}(E_{n}, \textbf P)\Big ]  \mathcal V^{\mathrm{out}}_{A_1,\{J'\}} }{\mathcal X^\dagger_{\{J\}} \Big [ \mathcal R_{\{J\}, \{J'\}}(E_{n}, \textbf P)\Big ]  \mathcal V^{\mathrm{out}}_{A_2,\{J'\}} }\,.
\end{align}

\section{One nontrivial example: $N+\mathcal{J}\rightarrow~(N\pi,N\eta,N\eta',\Sigma K,\Lambda K)$ with $|\textbf{d}|=1$ and $\ell\leq1$ \label{sec:NpiSPwave}}

Eq.~(\ref{eq:Fscdef}) is a concise definition of the finite volume function $F^{\mathrm{sc}}$. When considering a specific example it is more convenient to introduce an alternative representation first used in Ref.~\cite{Kim:2005gf},
\begin{eqnarray}
F^{\mathrm{sc}}_{alm_l,a'l'm'_l}(P,L)=\frac{iq^*_a}{8\pi E^*}\xi_a\left[\delta_{ll'}\delta_{m_lm_l'} +i\sum_{l'',m''}\frac{(4\pi)^{3/2}}{q^{*{l''}+1}_a}c_{l''m''}^{\mathbf{d}}(q^{*2}_a;{L})  \int d\Omega~Y^*_{ l,m_l}(\hat {\textbf k}^*_a) Y^*_{l'',m''}(\hat {\textbf k}^*_a) Y_{l',m_l'}(\hat {\textbf k}^*_a) \right],
\label{eq:FKSS}
\end{eqnarray}
where the $c^{\textbf{d}}_{lm}$  functions are defined via 
\begin{eqnarray}
\label{eq:clm}
c^\mathbf{d}_{lm}(k^{*2}_j; {L})
=\frac{\sqrt{4\pi}}{\gamma L^3}\left(\frac{2\pi}{L}\right)^{l-2}\mathcal{Z}^\mathbf{d}_{lm}[1;(k^*_j {L}/2\pi)^2],
\hspace{1cm}
\mathcal{Z}^\mathbf{d}_{lm}[s;x^2]
= \sum_{\mathbf r \in \mathcal{P}_{\mathbf{d}}}\frac{|\mathbf{r}|^lY_{lm}(\mathbf{r})}{(r^2-x^2)^s} \label{eq:clm} \,.
\end{eqnarray} 
The sum is performed over $\mathcal{P}_{\mathbf{d}}=\left\{\mathbf{r}\in \mathbb{R}^3\hspace{.1cm} | \hspace{.1cm}\mathbf{r}={\hat{\gamma}}^{-1}(\mathbf m-\alpha_a \mathbf d) \right\}$, $\textbf{m}$ is an integer triplet, $\mathbf d$ is the normalized boost vector $\mathbf d=\mathbf{P}L/2\pi$, $\alpha_a=\frac{1}{2}\left[1+\frac{m_{a,1}^2-m_{a,2}^2}{E^{*2}}\right]$, and $\hat{\gamma}^{-1}\textbf{x}\equiv{\gamma}^{-1}\textbf{x}_{||}+\textbf{x}_{\perp}$, with $\gamma = E/E^*$ and with $\textbf{x}_{||}(\textbf{x}_{\perp})$ denoting the $\textbf{x}$ component that is parallel  (perpendicular) to the total momentum, $\textbf P$. In Appendix~\ref{sec:assym}, we demonstrate how to generalize this to describe systems in asymmetric volumes and systems with twisted boundary conditions.

In Ref.~\cite{Briceno:2014uqa}, we demonstrated how to utilize the main result presented here, Eq.~(\ref{eq:onetotwomain}), for various systems with zero intrinsic spin. In order to illustrate the power of this generalization, we consider a simple scenario of $N\pi$ near threshold where contributions due to partial waves with $\ell\geq2$ can be ignored. Due the nonzero intrinsic spin, there are three partial waves we must consider $\delta_{\ell_J}=\{ \delta_{S_{1/2}},\delta_{P_{1/2}},\delta_{P_{3/2}}\}$. 
When the system is at rest, parity is a good quantum number and the S-wave and P-waves do not mix. When the system has nonzero total momentum, parity is no longer a good quantum number and consequently S-wave and P-waves can in fact mix. This mixing is explicitly illustrated in the finite-volume functions appearing in the quantization condition, Eq.~(\ref{eq:onetotwomain}), and the residue matrix, Eq.~(\ref{eq:Resfunc}). 

For example, when the system has a boost vector $|\textbf{d}|=1$, its symmetry group is $\text{Dic}_4$. Following the notation used in Ref.~\cite{Moore:2005dw}, the irreps that couple to half-integer spin systems are the $\mathbb{E}_1=S_{1/2}\oplus P_{1/2}\oplus P_{3/2}\oplus\cdots$ and $\mathbb{E}_3=P_{3/2}\oplus\cdots$. The finite-volume functions and scattering matrices corresponding to these irreps are
\begin{eqnarray}
\label{eq:KmatNpiG2}
 \text{Dic}_4 ~\mathbb{E}_1:~~~~~~
 %%%%%%%%%%%%%%%%%%
{F}^{}_{\mathbb{E}_1}&=&
\frac{q^*}{8\pi E^*}
\left(
\begin{array}{ccc}
\cot\phi^\textbf{d}_{00} +i& -\frac{\cot\phi^\textbf{d}_{10}}{\sqrt{3}} & \sqrt{\frac{2}{3}}\cot\phi^\textbf{d}_{10} \\
 -\frac{\cot\phi^\textbf{d}_{10}}{\sqrt{3}} &\cot\phi^\textbf{d}_{00} +i& -\sqrt{\frac{2}{5}}\cot\phi^\textbf{d}_{20} \\
 \sqrt{\frac{2}{3}}\cot\phi^\textbf{d}_{10} & -\sqrt{\frac{2}{5}}\cot\phi^\textbf{d}_{20} &\cot\phi^\textbf{d}_{00}+\frac{\cot\phi^\textbf{d}_{20}}{\sqrt{5}}+i \\
\end{array}
\right)
,\\
\mathcal{M}_{\mathbb{E}_1}&=&
\frac{8\pi E^*}{q^*}
\left(
\begin{array}{ccc}
[\cot\delta_{S_{\frac{1}{2}}}-i]^{-1}  & 0 & 0 \\
0&[\cot\delta_{P_{\frac{1}{2}}}-i]^{-1} &0 \\
0&0&[\cot\delta_{P_{\frac{3}{2}}}-i]^{-1}  \\
\end{array}
\right),
\label{eq:FVG2Dic4}
\\\label{eq:KmatNpiG3}
 \text{Dic}_4 ~\mathbb{E}_3:~~~~~~
 %%%%%%%%%%%%%%%%%%
{F}^{}_{\mathbb{E}_3}&=&
\frac{q^*}{8\pi E^*}
\left(\cot\phi^\textbf{d}_{00}-\frac{\cot\phi^\textbf{d}_{20}}{\sqrt{5}} +i\right)
,~~~~
\mathcal{M}_{\mathbb{E}_3}=
\frac{8\pi E^*}{q^*}
\frac{1}{\cot\delta_{P_{\frac{3}{2}}}-i},
\label{eq:FVG3Dic4}
\end{eqnarray} 
where the pseudophases, $\phi^\textbf{d}_{lm}$, are defined via
\begin{eqnarray}
q^*_{\Lambda,n}\cot\phi^\textbf{d}_{lm}=-\frac{4\pi}{q^{*l}_{\Lambda,n}} c^\mathbf{d}_{lm}(q^{*2}_{\Lambda,n}; {L}).
\label{eq:philm}
\end{eqnarray} 

 Determining these matrices as functions of $E$ and $L$ and substituting into Eq.~(\ref{eq:Resfunc}) gives a three-by-three matrix, $\mathcal R_{\mathbb E_1}(E_{\mathbb E_1, n}, \textbf P)$, and a single value, $\mathcal R_{\mathbb E_3}(E_{\mathbb E_3, n}, \textbf P)$, for each energy level in the finite-volume spectra of the indicated irreps. The single value, $\mathcal R_{\mathbb E_3}(E_{\mathbb E_3, n}, \textbf P)$, gives a Lellouch-L\"uscher like proportionality, relating the finite-volume matrix element of the $n$th state to the transition amplitude
\begin{equation}
\big \vert \langle E'_{ n}, \textbf P', L ,\mathbb E_3 \big \vert \widetilde {\mathcal J}_A(0, \textbf P - \textbf P') \vert E_{0, \textbf P}, \textbf P,L ,1\rangle \big \vert  = \frac{1}{\sqrt{2E_{0,\textbf P}}} \sqrt{ \big \vert \mathcal  R_{\mathbb E_3}(E_{ n}', \textbf P') \big \vert} \ \big \vert \mathcal H^{\mathrm{out}}_{A, P_{\frac32}} \big \vert  \,,
\end{equation}
where
\begin{equation}
\big[ \mathcal H ^{\mathrm{out}}_{A, P_{\frac32}}(E_{0,\textbf P}, \textbf P; E_{n}', \textbf P') \big] \ (2 \pi)^3  \delta^3(\textbf P - \textbf P' - \textbf Q) \equiv  \langle E'_{n}, \textbf P',N, \pi, P_{\frac32}, \mathrm{out} \vert \widetilde {\mathcal J}_A(0, \textbf Q)\vert E_{0,\textbf P} , \textbf P, N   \rangle \,.
\end{equation}
In this case one can easily simplify the form of $\mathcal R_{\mathbb E_3}$
\begin{align}
\mathcal R_{\mathbb E_3}(E_{n}, \textbf P) & =   \left [ \frac{\partial}{\partial E} \left( F^{-1}(P,L) + \mathcal M(P) \right ) \right ]_{E=E_n}^{-1} = -  \left [ \mathcal M^2(P) \frac{\partial}{\partial E} \left( F(P,L) + \mathcal M^{-1}(P) \right ) \right ]_{E=E_n}^{-1} \,, \\
& = -\frac{q^*}{8 \pi E^*}   \left [ \sin^2\! \delta_{P_{\frac{3}{2}}}\ e^{ 2i \delta_{P_{{3}/{2}}}}\  \frac{\partial}{\partial E} \left( \cot\phi^\textbf{d}_{00}-\frac{\cot\phi^\textbf{d}_{20}}{\sqrt{5}} + \cot\delta_{P_{\frac{3}{2}}} \right ) \right ]_{E=E_n}^{-1}    \,, \\
& = \frac{q^*}{8 \pi E^*} e^{-2i \delta_{P_{{3}/{2}}}}  \left [  \frac{\partial}{\partial E} \left( \phi^\textbf{d}_{\mathbb E_3} + \delta_{P_{\frac{3}{2}}} \right ) \right ]_{E=E_n}^{-1}    \,,
 \end{align}
where we have introduced
\begin{equation}
\cot \phi^\textbf{d}_{\mathbb E_3} \equiv \cot\phi^\textbf{d}_{00}-\frac{\cot\phi^\textbf{d}_{20}}{\sqrt{5}} \,.
\end{equation}
Note that the phase appearing in $\mathcal R_{\mathbb E_3}$ is exactly that needed to cancel the phases from $\mathcal H^{\mathrm{in}}$ and $\mathcal H^{\mathrm{out}}$ so that the right-hand side is pure real.

For the three-by-three matrix, $\mathcal R_{\mathbb E_1}$, there is no straightforward reduction of the general relation, Eq.~(\ref{eq:onetotwomain}). In this case the result relates the finite volume matrix elements of states that transform in $\mathbb E_1$ to linear combinations of three transition amplitudes
\begin{equation}
\big[ \mathcal H^{\mathrm{out}}_{A, \mathbb E_1}(E_{0,\textbf P}, \textbf P; E_{n}', \textbf P') \big] \ (2 \pi)^3  \delta^3(\textbf P - \textbf P' - \textbf Q) \equiv \begin{pmatrix} \langle E'_{n}, \textbf P',N, \pi, S_{\frac12}, \mathrm{out} \vert \widetilde {\mathcal J}_A(0, \textbf Q)\vert E_{0,\textbf P} , \textbf P, N   \rangle \\
\langle E'_{n}, \textbf P',N, \pi, P_{\frac12}, \mathrm{out} \vert \widetilde {\mathcal J}_A(0, \textbf Q)\vert E_{0,\textbf P} , \textbf P, N   \rangle \\
\langle E'_{n}, \textbf P',N, \pi, P_{\frac32}, \mathrm{out} \vert \widetilde {\mathcal J}_A(0, \textbf Q)\vert E_{0,\textbf P} , \textbf P, N   \rangle \end{pmatrix} \,.
\end{equation}

For physical or nearly physical quark masses, the three-particle ($N\pi\pi$) threshold resides close to the $N\pi$ threshold. For unphysically heavy quark masses, the three-particle threshold is pushed up and other two-particle thresholds ($N\eta,N\eta',\Sigma K,\Lambda K$) approach the $N\pi$ threshold. In the SU(3) flavor limit all of these thresholds overlap. As we have discussed, incorporating more than one two-body channel amounts to upgrading the angular-momentum matrices to matrices in the product space of flavor and angular momentum.

 \section{Conclusion}
In this work we present a relation between finite-volume matrix elements and $\textbf{1}\rightarrow\textbf{2}$ as well as $\textbf{0}\rightarrow\textbf{2}$ transition amplitudes in the presence of an external current. The result is exact up to exponentially suppressed volume corrections and is nonpertubative in the strong dynamics. The result presented here is the most general of its kind, and holds when the individual particles have arbitrary spin and the final state is composed of any number of strongly coupled two-body channels. Furthermore, the result is independent of the nature of the external current. It may therefore be implemented for determining electroweak transition amplitudes as well as BSM processes. 

{ 
To study transition amplitudes involving two particles in the initial or final state, one must first constrain the scattering amplitude. This can be done by determining the finite-volume spectrum and then applying Eq.~(\ref{eq:QC})/(\ref{eq:QCinSecFour}) as has been done, for example, in Refs.~\cite{Dudek:2012xn, Dudek:2014qha, Wilson:2014cna}. Once the phase shift and mixing angles have been parametrized, one may proceed to take derivatives of these to determine the residue function $\mathcal R$, Eq.~(\ref{eq:Rintro})/(\ref{eq:Resfunc}), which allows one to relate finite-volume matrix elements with transition amplitudes. In general, in order to determine transition processes involving resonances, it is necessary to determine not just the matrix element of the ground state but also excited states. Recently, Ref.~\cite{Shultz:2015pfa} has demonstrated an efficient way to determine excited state matrix-elements using the distillation framework. Having constrained both the residue function and the finite volume matrix elements for $\textbf{1} \rightarrow \textbf{2}$ and $\textbf{0} \rightarrow \textbf{2}$ processes, one may proceed to access the corresponding transition amplitudes using Eqs.~(\ref{eq:matJaieps})/(\ref{eq:onetotwomain}) and (\ref{eq:matJ0to1})/(\ref{eq:zerototwomain}), respectively.
}

\noindent
\subsection*{Acknowledgments}
RB acknowledges support from the U.S. Department of Energy contract DE-AC05-06OR23177, under which Jefferson Science Associates, LLC, manages and operates the Jefferson Lab. The authors would like to acknowledge and thank Jozef Dudek, Robert Edwards, Stefan Meinel, Stephen Sharpe, Christian Shultz, Christopher Thomas, Andr\'e Walker-Loud and David Wilson for useful discussion.   
\noindent

%%%%%%%%%%%%%%%%%%%%%%%%%%%%%%%%%%%%%%%%%%%%%%%%%%
%%%%%%%%%%%%%%%%%%     APPENDICES   %%%%%%%%%%%%%%%%%%%%%%
%%%%%%%%%%%%%%%%%%%%%%%%%%%%%%%%%%%%%%%%%%%%%%%%%%

\appendix

\section{Free limit \label{sec:free_limit}}

To give further insight to Eq.~(\ref{eq:onetotwomain}), here we  first consider the result near two-particle production threshold and then in the case where the two outgoing particles are non-interacting. In the former case, we set all phase shifts to zero with the exception of the S-wave. In this limit, the ratio between the transition amplitude and finite volume matrix element simplifies to
\begin{eqnarray}
\frac{|\mathcal{H}_{S,n}|^2}
{|\langle E'_{n}, \textbf{P}',L|\widetilde{\mathcal{J}}(0,\textbf{P}-\textbf{P}')| E_{0,\textbf P}, \textbf{P},  L, 1\rangle|^2}=
\frac{16\pi E_{0, \textbf P}~E'^{*}_{n}}
{\xi\mathcal{N}_i\mathcal{N}_f~q'^{*}_{n}}
\left.~\frac{\partial (\delta_S+\phi^\textbf{d}_{00})}
{\partial E'}\right|
_{ E'=E'_{n}},
\label{eq:LLfactor}
\end{eqnarray}
 where, for clarity, we have explicitly included the normalization factors of the initial ($\mathcal{N}_i$) and final ($\mathcal{N}_f$) finite-volume states. 
Next, using Eq.~(\ref{eq:philm}), one obtains a simple result in the limit in which the final two particles do not interact,
\begin{align}
\frac{|\mathcal{H}_{S,n}|^2}
{|\langle E'_{n}, \textbf{P}',L|\widetilde{\mathcal{J}}(0,\textbf{P}-\textbf{P}')| E_{0,\textbf P}, \textbf{P},  L, 1\rangle|^2}
&\approx
\frac{2E_{0,\textbf P}}{\mathcal{N}_i\mathcal{N}_f}~\mathcal{R}_{free}^{-1}
\equiv \frac{2E_{0,\textbf P}}{\xi\mathcal{N}_i\mathcal{N}_f~}\frac{E'^2_{n}}{\nu_{n}} L^3,
\label{eq:free}
\end{align}
where $\nu_{n}$ is the degeneracy of the $n$th state, equivalently the number of vectors $\textbf n \in \mathbb Z^3$ such that $\textbf n^2 = n$. In general, one expects $\mathcal{R}/\mathcal{R}_{free}\sim\mathcal{O}(1)$ for weakly interacting systems. In arriving at this result, we have used the fact that the phase shift, and thus also its first derivative, vanishes, as well as the identity $\partial \phi^\textbf{d}_{00}=\cos^2 \phi^\textbf{d}_{00}~\partial \tan\phi^\textbf{d}_{00}$.

%%%%%%%%%%%%%%%%%%%%%%%%%%%%%%%%%%%%%%%%%%%
\begin{figure*}[t]
\begin{center}
\subfigure[]{
\label{fig:resonance_M}
\includegraphics[scale=.6]{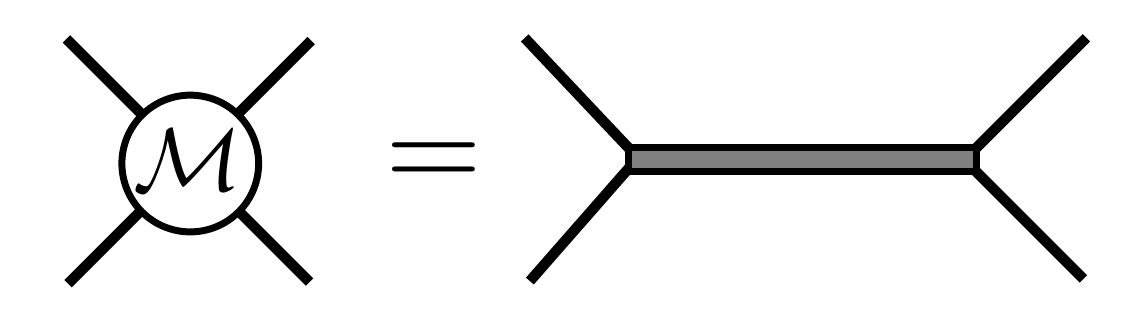}}\hspace{1cm}
\subfigure[]{
\label{fig:resonance_H}
\includegraphics[scale=0.6]{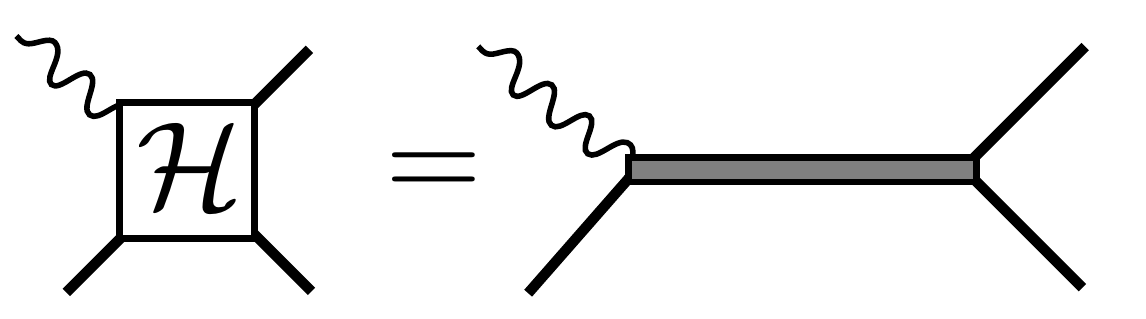}}
\caption{(a) In the vicinity of a resonance, the two-body scattering amplitude can be obtained by mediating the two-body systems using a fully dressed auxiliary field. (b) Using this same auxiliary field, one can also obtain an expression for the $\textbf{1}\rightarrow\textbf{2}$ transition amplitude in the presence of an external field, Eq.~(\ref{eq:resonance_H}).  }\label{fig:resonanceamp}
\end{center}
\end{figure*}
%%%%%%%%%%%%%%%%%%%%%%%%%%%%%%%%%%%%%%%%%%%
\section{Narrow-width approximation~\label{sec:narrow_width}}

Another interesting limit, which was previously considered in Ref.~\cite{Agadjanov:2014kha}, is the narrow width approximation. Here the exact nature of the resonance is not relevant, and thus we leave the partial wave, $\ell$, unspecified. In the narrow-width limit, the two-body resonance approaches a bound state. A resonance corresponds to a pole in the complex plane, with the imaginary part of the pole location proportional to the resonance width. If we send the width towards zero, then the pole approaches the real axis and the state becomes bound. In order to consider this limit, we investigate the behavior of the scattering amplitude and transition amplitudes near a resonance. We begin by considering the two-particle scattering amplitude at energies near the resonance pole. Using a \emph{Breit-Wigner inspired} parametrization, we write the scattering amplitude as~[see Fig.~\ref{fig:resonance_M}]
\begin{eqnarray}
\mathcal{M}_{\ell}
=\frac{G_{\textbf{2}\rightarrow R}^2(E^*)}{m_R^2-E^{*2}-i E^*\Gamma(E^*)},
\label{eq:resonance_M}
\end{eqnarray}
where $G_{\textbf{2}\rightarrow R}^2(E^*)$ is a generic function that parametrizes the coupling of the two-particle state to the intermediate resonance, $m_R$ is the real part of the resonance pole location and $\Gamma(E^*)$ is its energy dependent decay width. From Eq.~(\ref{Smatrix}) above, we know that the scattering amplitude must satisfy
\begin{eqnarray}
\mathcal{M}_{\ell}
=\frac{8\pi E^*}{\xi q^*}\frac{1}{\cot\delta_\ell-i}.
\end{eqnarray}
Equating these two expressions we reach the following relations for $G_{\textbf{2}\rightarrow R}$, $\Gamma$ and the scattering phase shift,
\begin{align}
G_{\textbf{2}\rightarrow R}(E^*)
=\frac{8\pi E^{*2}\Gamma(E^*)}{\xi q^*}
,
~~~~
\tan \delta(E^*)=\frac{E^*~\Gamma(E^*)}{(m_R^2-E^{*2})} \,.
\label{eq:BWps}
\end{align}

Similarly, we define the transition amplitude in the vicinity of the resonance as~[see Fig.~\ref{fig:resonance_H}]
\begin{align}
\mathcal{H}_{\ell}
&=
\frac{
F^{\textbf{1}\rightarrow R}_{c}(E^*,Q^2)~
G_{\textbf{2}\rightarrow R}(E^*)
}{m_R^2-E^{*2}-i E^*\Gamma(E^*)} \,,
\nn\\
&={F^{\textbf{1}\rightarrow R}_{c}(E^*,Q^2)}\frac{1}{m_R^2-E^{*2}-i E^*\Gamma(E^*)}~\sqrt{\frac{8\pi E^{*2}\Gamma(E^*)}{\xi q^*}},
\label{eq:resonance_H}
\end{align}
where $F^{\textbf{1}\rightarrow R}_{c}(E^*,Q^2)$ is a smooth function of both $E^*$ and $Q^2=(P-P')^2$. Note that although the numerator of the scattering amplitude in the presence of a resonance is proportional to $\Gamma(E^*)$, the transition amplitude is proportional to $\sqrt{\Gamma(E^*)}$.

Now we assume we are in the vicinity of a single resonance state. Using the parametrization of the phase shift given above and defining $\Gamma_R\equiv\Gamma(m_R)$, one finds the derivative of the phase shift as it approaches  $\pi/2$ to be 
\begin{equation}
\left. 
\frac{\partial}{\partial E} \delta(E^*) \right|_{E^*=m_R}=
\left. 
-\sin^2\delta\frac{\partial}{\partial E} \cot\delta(E^*)\right|_{E^*=m_R}
= 2\frac{E_R}{\Gamma_Rm_R} \left[1 + \mathcal O(\Gamma_R/m_R) \right ] \,,
\end{equation}
where $E_R\equiv\sqrt{\textbf{P}^2+m_R^2}$. As the width goes to zero the derivative of the phase shift divergences as expected. Ignoring the contribution from the derivative of the pseudophases, which are finite away from free-particle poles, one obtains the following result for the right hand side of Eq.~(\ref{eq:LLfactor}) near a narrow-width resonance
\begin{equation}
\frac{|\mathcal{H}_{\ell}|^2}
{|\langle E_{R}, \textbf{P}',L|\widetilde{\mathcal{J}}(0,\textbf{P}-\textbf{P}')| E_{0,\textbf P}, \textbf{P},  L, 1\rangle|^2}
=
\frac{2E_{0,\textbf P}}{\xi\mathcal{N}_i\mathcal{N}_f}
\frac{16 \pi E_R}{q^{*}_{R} \Gamma_R}  \left[1 + \mathcal O(\Gamma_R/m_R) \right ] \,,
\label{eq:LLfactorP2}
\end{equation}

From Eq.~(\ref{eq:resonance_H}) we find that near the resonance mass, the transition amplitude divergences inversely proportional to the square root of the resonance width
\begin{eqnarray}
\mathcal{H}_{\ell}
=
{F^{\textbf{1}\rightarrow R}_{c}(m_R,Q^2)}
~\sqrt{\frac{8\pi }{\xi q^*_R\Gamma_R}} \left[1 + \mathcal O(\Gamma_R/m_R) \right ],
\end{eqnarray} 

In this limit, one finds that the finite-volume matrix element is equal to the infinite volume $\textbf{1}\rightarrow R$ transition amplitude up to the standard normalization of the states,
\begin{equation}
|\langle E_{R}, \textbf{P}',L|\widetilde{\mathcal{J}}(0,\textbf{P}-\textbf{P}')| E_{0,\textbf P}, \textbf{P},  L, 1\rangle|^2
=
{|F^{\textbf{1}\rightarrow R}_{c}(m_R,Q^2)|^2} 
\frac{\mathcal{N}_i\mathcal{N}_f}
{2E_{0,\textbf P}2 E_R} \left[ 1 + \mathcal O(\Gamma_R/m_R) \right ]\,.
\label{eq:pi_to_Rho}
\end{equation}
 It is important to emphasize that this approximation only holds when the resonance is very narrow ($\Gamma_R/m_R\ll 1$) and the energy level determined corresponds to the resonances mass, up to small corrections that scale with the width. To reliably asses the validity of this approximation, one must first determine the phase shift as a function of the energy using Eq.~(\ref{eq:QCinSecFour}) as done, for example, in Ref.~\cite{Dudek:2012xn} for the $\rho$-resonance. Since a zero width resonance is equivalent to a two-body bound state, this discussion applies also for bound states, when the width is exactly zero.

%%%%%%%%%%%%%%%%%%%%%%%%%%%%%%%%%%%%%%%%%%%%%%%%%%%%%%%%%%%%%%

\section{Asymmetric boxes with twisted boundary conditions \label{sec:assym}}
Although the vast majority of present day calculations are performed in symmetric volumes with periodic boundary conditions~\cite{Bedaque:2004kc}, the formalism presented here can be generalized to volumes that are arbitrary rectangular prisms with twisted boundary conditions. Following the notation used in Refs.~\cite{Briceno:2014oea, Briceno:2014uqa}, we define $\bm{\phi}_{a,1}$ and $\bm{\phi}_{a,2}$ to be the three-dimensional phases of the first and second particle, respectively, in the $a$th channel. We also define $L$ to be the spatial extent of the z-axis and $\eta_i$ to be defined such that $L_x=\eta_xL$ and  $L_y=\eta_yL$. Furthermore, let the vector $\tilde{\bm{\chi}}=(\chi_x/\eta_x,\chi_y/\eta_y,\chi_z)$. With this, we can express the generalization of the $c^\mathbf{d}_{lm}$ functions in Eq.~(\ref{eq:clm}),
\begin{eqnarray}
c^{\textbf{d},\bm{\phi}_{a,1},\bm{\phi}_{a,2}}_{lm}(k^{*2};{L};\eta_x,\eta_y)
\ &=&\ \frac{\sqrt{4\pi}}{\eta_x\eta_y\gamma {L}^3}\left(\frac{2\pi}{{L}}\right)^{l-2}\times\mathcal{Z}^{\mathbf{d},\bm{\phi}_{a,1},\bm{\phi}_{a,2}}_{lm}[1;(k^*{L}/2\pi)^2;\eta_x,\eta_y],~~
\label{clmasymTBCs}\\
\mathcal{Z}^{\mathbf{d},\bm{\phi}_{a,1},\bm{\phi}_{a,2}}_{lm}[s;x^2;\eta_x,\eta_y]
\ &=&\ \sum_{\mathbf r \in \mathcal{P}_{\mathbf{d};\eta_x,\eta_y}^{\bm{\phi}_1,\bm{\phi}_2;}}
\frac{ |{\bf r}|^l \ Y_{lm}(\mathbf{r})}{(\mathbf{r}^2-x^2)^s},~~~
\label{ZlmasymTBCs}
\end{eqnarray}
where $ \mathcal{P}_{\mathbf{d};\eta_x,\eta_y}^{\bm{\phi}_1,\bm{\phi}_2}=\left\{\mathbf{r}\in \textbf{R}^3\hspace{.1cm} | \hspace{.1cm}\mathbf{r}={\hat{\gamma}}^{-1}(\tilde{\mathbf m}-\alpha_a \tilde{\mathbf d} +\frac{\tilde{\bm{\Delta}}^{(a)}}{2\pi})\right\}$, with $\textbf{m}$ a triplet integer, $\tilde{\bm{\Delta}}^{(a)}=-({\alpha}_{a}-\frac{1}{2})(\tilde{\bm{\phi}}_{a,1}+\tilde{\bm{\phi}}_{a,2})+\frac{1}{2}(\tilde{\bm{\phi}}_{a,1}-\tilde{\bm{\phi}}_{a,2})$, $\tilde{\mathbf d}={\mathbf P}L/2\pi$, and all other quantities defined after Eq.~(\ref{eq:clm}).  

 By replacing the factors of $F$ in all expressions with this generalization, and additionally replacing all factors of $L^3$ with $\eta_x \eta_y L^3$, one reaches extensions of all results that are valid for arbitrary rectangular prisms with arbitrary twist. As discussed in Ref.~\cite{Briceno:2013hya}, in the presence of general twisted boundary conditions, the symmetry of the system is further reduced. Depending on the twists chosen, one may need to rederive the $F$-functions appearing in Eqs.~(\ref{eq:KmatNpiG2}) and (\ref{eq:FVG3Dic4}). We point the reader to Refs.~\cite{Briceno:2014oea, Briceno:2014uqa, Briceno:2013hya} for further details.  
%%%%%%%%%%%%%%%%%%%%%%%%%%%%%%%%%%%%%%%%%%%%%%%%%%%%%%%%%%%%%%

\bibliographystyle{apsrev} %%% physical review
\bibliography{bibi} %%% ref.bib file

\begin{thebibliography}{57}
\expandafter\ifx\csname natexlab\endcsname\relax\def\natexlab#1{#1}\fi
\expandafter\ifx\csname bibnamefont\endcsname\relax
  \def\bibnamefont#1{#1}\fi
\expandafter\ifx\csname bibfnamefont\endcsname\relax
  \def\bibfnamefont#1{#1}\fi
\expandafter\ifx\csname citenamefont\endcsname\relax
  \def\citenamefont#1{#1}\fi
\expandafter\ifx\csname url\endcsname\relax
  \def\url#1{\texttt{#1}}\fi
\expandafter\ifx\csname urlprefix\endcsname\relax\def\urlprefix{URL }\fi
\providecommand{\bibinfo}[2]{#2}
\providecommand{\eprint}[2][]{\url{#2}}

\bibitem[{\citenamefont{AlekSejevs et~al.}(2013)}]{AlekSejevs:2013mkl}
\bibinfo{author}{\bibfnamefont{A.}~\bibnamefont{AlekSejevs}}
  \bibnamefont{et~al.} (\bibinfo{collaboration}{GlueX Collaboration})
  (\bibinfo{year}{2013}), \eprint{1305.1523}.

\bibitem[{\citenamefont{Shultz et~al.}(2015)\citenamefont{Shultz, Dudek, and
  Edwards}}]{Shultz:2015pfa}
\bibinfo{author}{\bibfnamefont{C.~J.} \bibnamefont{Shultz}},
  \bibinfo{author}{\bibfnamefont{J.~J.} \bibnamefont{Dudek}}, \bibnamefont{and}
  \bibinfo{author}{\bibfnamefont{R.~G.} \bibnamefont{Edwards}}
  (\bibinfo{year}{2015}), \eprint{1501.07457}.

\bibitem[{\citenamefont{Wei et~al.}(2009)}]{Wei:2009zv}
\bibinfo{author}{\bibfnamefont{J.-T.} \bibnamefont{Wei}} \bibnamefont{et~al.}
  (\bibinfo{collaboration}{BELLE Collaboration}),
  \bibinfo{journal}{Phys.Rev.Lett.} \textbf{\bibinfo{volume}{103}},
  \bibinfo{pages}{171801} (\bibinfo{year}{2009}), \eprint{0904.0770}.

\bibitem[{\citenamefont{Aaltonen et~al.}(2012)}]{Aaltonen:2011ja}
\bibinfo{author}{\bibfnamefont{T.}~\bibnamefont{Aaltonen}} \bibnamefont{et~al.}
  (\bibinfo{collaboration}{CDF Collaboration}),
  \bibinfo{journal}{Phys.Rev.Lett.} \textbf{\bibinfo{volume}{108}},
  \bibinfo{pages}{081807} (\bibinfo{year}{2012}), \eprint{1108.0695}.

\bibitem[{\citenamefont{Lees et~al.}(2012)}]{Lees:2012tva}
\bibinfo{author}{\bibfnamefont{J.}~\bibnamefont{Lees}} \bibnamefont{et~al.}
  (\bibinfo{collaboration}{BaBar Collaboration}), \bibinfo{journal}{Phys.Rev.}
  \textbf{\bibinfo{volume}{D86}}, \bibinfo{pages}{032012}
  (\bibinfo{year}{2012}), \eprint{1204.3933}.

\bibitem[{\citenamefont{Aaij et~al.}(2013{\natexlab{a}})}]{Aaij:2013iag}
\bibinfo{author}{\bibfnamefont{R.}~\bibnamefont{Aaij}} \bibnamefont{et~al.}
  (\bibinfo{collaboration}{LHCb Collaboration}), \bibinfo{journal}{JHEP}
  \textbf{\bibinfo{volume}{1308}}, \bibinfo{pages}{131}
  (\bibinfo{year}{2013}{\natexlab{a}}), \eprint{1304.6325}.

\bibitem[{\citenamefont{Aaij et~al.}(2013{\natexlab{b}})}]{Aaij:2013qta}
\bibinfo{author}{\bibfnamefont{R.}~\bibnamefont{Aaij}} \bibnamefont{et~al.}
  (\bibinfo{collaboration}{LHCb collaboration}),
  \bibinfo{journal}{Phys.Rev.Lett.} \textbf{\bibinfo{volume}{111}},
  \bibinfo{pages}{191801} (\bibinfo{year}{2013}{\natexlab{b}}),
  \eprint{1308.1707}.

\bibitem[{\citenamefont{Bobeth et~al.}(2013)\citenamefont{Bobeth, Hiller, and
  van Dyk}}]{Bobeth:2012vn}
\bibinfo{author}{\bibfnamefont{C.}~\bibnamefont{Bobeth}},
  \bibinfo{author}{\bibfnamefont{G.}~\bibnamefont{Hiller}}, \bibnamefont{and}
  \bibinfo{author}{\bibfnamefont{D.}~\bibnamefont{van Dyk}},
  \bibinfo{journal}{Phys.Rev.} \textbf{\bibinfo{volume}{D87}},
  \bibinfo{pages}{034016} (\bibinfo{year}{2013}), \eprint{1212.2321}.

\bibitem[{\citenamefont{Descotes-Genon
  et~al.}(2013)\citenamefont{Descotes-Genon, Matias, and
  Virto}}]{Descotes-Genon:2013wba}
\bibinfo{author}{\bibfnamefont{S.}~\bibnamefont{Descotes-Genon}},
  \bibinfo{author}{\bibfnamefont{J.}~\bibnamefont{Matias}}, \bibnamefont{and}
  \bibinfo{author}{\bibfnamefont{J.}~\bibnamefont{Virto}},
  \bibinfo{journal}{Phys.Rev.} \textbf{\bibinfo{volume}{D88}},
  \bibinfo{pages}{074002} (\bibinfo{year}{2013}), \eprint{1307.5683}.

\bibitem[{\citenamefont{Hambrock et~al.}(2014)\citenamefont{Hambrock, Hiller,
  Schacht, and Zwicky}}]{Hambrock:2013zya}
\bibinfo{author}{\bibfnamefont{C.}~\bibnamefont{Hambrock}},
  \bibinfo{author}{\bibfnamefont{G.}~\bibnamefont{Hiller}},
  \bibinfo{author}{\bibfnamefont{S.}~\bibnamefont{Schacht}}, \bibnamefont{and}
  \bibinfo{author}{\bibfnamefont{R.}~\bibnamefont{Zwicky}},
  \bibinfo{journal}{Phys.Rev.} \textbf{\bibinfo{volume}{D89}},
  \bibinfo{pages}{074014} (\bibinfo{year}{2014}), \eprint{1308.4379}.

\bibitem[{\citenamefont{Beaujean et~al.}(2014)\citenamefont{Beaujean, Bobeth,
  and van Dyk}}]{Beaujean:2013soa}
\bibinfo{author}{\bibfnamefont{F.}~\bibnamefont{Beaujean}},
  \bibinfo{author}{\bibfnamefont{C.}~\bibnamefont{Bobeth}}, \bibnamefont{and}
  \bibinfo{author}{\bibfnamefont{D.}~\bibnamefont{van Dyk}},
  \bibinfo{journal}{Eur.Phys.J.} \textbf{\bibinfo{volume}{C74}},
  \bibinfo{pages}{2897} (\bibinfo{year}{2014}), \eprint{1310.2478}.

\bibitem[{\citenamefont{Horgan et~al.}(2014{\natexlab{a}})\citenamefont{Horgan,
  Liu, Meinel, and Wingate}}]{Horgan:2013hoa}
\bibinfo{author}{\bibfnamefont{R.~R.} \bibnamefont{Horgan}},
  \bibinfo{author}{\bibfnamefont{Z.}~\bibnamefont{Liu}},
  \bibinfo{author}{\bibfnamefont{S.}~\bibnamefont{Meinel}}, \bibnamefont{and}
  \bibinfo{author}{\bibfnamefont{M.}~\bibnamefont{Wingate}},
  \bibinfo{journal}{Phys.Rev.} \textbf{\bibinfo{volume}{D89}},
  \bibinfo{pages}{094501} (\bibinfo{year}{2014}{\natexlab{a}}),
  \eprint{1310.3722}.

\bibitem[{\citenamefont{Horgan et~al.}(2014{\natexlab{b}})\citenamefont{Horgan,
  Liu, Meinel, and Wingate}}]{Horgan:2013pva}
\bibinfo{author}{\bibfnamefont{R.~R.} \bibnamefont{Horgan}},
  \bibinfo{author}{\bibfnamefont{Z.}~\bibnamefont{Liu}},
  \bibinfo{author}{\bibfnamefont{S.}~\bibnamefont{Meinel}}, \bibnamefont{and}
  \bibinfo{author}{\bibfnamefont{M.}~\bibnamefont{Wingate}},
  \bibinfo{journal}{Phys.Rev.Lett.} \textbf{\bibinfo{volume}{112}},
  \bibinfo{pages}{212003} (\bibinfo{year}{2014}{\natexlab{b}}),
  \eprint{1310.3887}.

\bibitem[{\citenamefont{Bouchard}(2015)}]{Bouchard:2015pda}
\bibinfo{author}{\bibfnamefont{C.}~\bibnamefont{Bouchard}}
  (\bibinfo{year}{2015}), \eprint{1501.03204}.

\bibitem[{\citenamefont{Bailey et~al.}(2014)}]{Bailey:2014fpx}
\bibinfo{author}{\bibfnamefont{J.}~\bibnamefont{Bailey}} \bibnamefont{et~al.}
  (\bibinfo{collaboration}{Fermilab Lattice Collaboration, MILC Collaboration})
  (\bibinfo{year}{2014}), \eprint{1411.6038}.

\bibitem[{\citenamefont{Bouchard et~al.}(2014)\citenamefont{Bouchard, Lepage,
  Monahan, Na, and Shigemitsu}}]{Bouchard:2014ypa}
\bibinfo{author}{\bibfnamefont{C.}~\bibnamefont{Bouchard}},
  \bibinfo{author}{\bibfnamefont{G.~P.} \bibnamefont{Lepage}},
  \bibinfo{author}{\bibfnamefont{C.}~\bibnamefont{Monahan}},
  \bibinfo{author}{\bibfnamefont{H.}~\bibnamefont{Na}}, \bibnamefont{and}
  \bibinfo{author}{\bibfnamefont{J.}~\bibnamefont{Shigemitsu}},
  \bibinfo{journal}{Phys.Rev.} \textbf{\bibinfo{volume}{D90}},
  \bibinfo{pages}{054506} (\bibinfo{year}{2014}), \eprint{1406.2279}.

\bibitem[{\citenamefont{Brice\~no
  et~al.}(2014{\natexlab{a}})\citenamefont{Brice\~no, Davoudi, and
  Luu}}]{Briceno:2014tqa}
\bibinfo{author}{\bibfnamefont{R.~A.} \bibnamefont{Brice\~no}},
  \bibinfo{author}{\bibfnamefont{Z.}~\bibnamefont{Davoudi}}, \bibnamefont{and}
  \bibinfo{author}{\bibfnamefont{T.~C.} \bibnamefont{Luu}}
  (\bibinfo{year}{2014}{\natexlab{a}}), \eprint{1406.5673}.

\bibitem[{\citenamefont{Brice\~no}(2014{\natexlab{a}})}]{Briceno:2014pka}
\bibinfo{author}{\bibfnamefont{R.~A.} \bibnamefont{Brice\~no}}
  (\bibinfo{year}{2014}{\natexlab{a}}), \eprint{1411.6944}.

\bibitem[{\citenamefont{Beane et~al.}(2014)\citenamefont{Beane, Detmold,
  Orginos, and Savage}}]{Beane:2014oea}
\bibinfo{author}{\bibfnamefont{S.~R.} \bibnamefont{Beane}},
  \bibinfo{author}{\bibfnamefont{W.}~\bibnamefont{Detmold}},
  \bibinfo{author}{\bibfnamefont{K.}~\bibnamefont{Orginos}}, \bibnamefont{and}
  \bibinfo{author}{\bibfnamefont{M.~J.} \bibnamefont{Savage}}
  (\bibinfo{year}{2014}), \eprint{1410.2937}.

\bibitem[{\citenamefont{Lellouch and L\"uscher}(2001)}]{Lellouch:2000pv}
\bibinfo{author}{\bibfnamefont{L.}~\bibnamefont{Lellouch}} \bibnamefont{and}
  \bibinfo{author}{\bibfnamefont{M.}~\bibnamefont{L\"uscher}},
  \bibinfo{journal}{Commun.Math.Phys.} \textbf{\bibinfo{volume}{219}},
  \bibinfo{pages}{31} (\bibinfo{year}{2001}), \eprint{hep-lat/0003023}.

\bibitem[{\citenamefont{Lin et~al.}(2001)\citenamefont{Lin, Martinelli,
  Sachrajda, and Testa}}]{Lin:2001ek}
\bibinfo{author}{\bibfnamefont{C.~D.} \bibnamefont{Lin}},
  \bibinfo{author}{\bibfnamefont{G.}~\bibnamefont{Martinelli}},
  \bibinfo{author}{\bibfnamefont{C.~T.} \bibnamefont{Sachrajda}},
  \bibnamefont{and} \bibinfo{author}{\bibfnamefont{M.}~\bibnamefont{Testa}},
  \bibinfo{journal}{Nucl.Phys.} \textbf{\bibinfo{volume}{B619}},
  \bibinfo{pages}{467} (\bibinfo{year}{2001}), \eprint{hep-lat/0104006}.

\bibitem[{\citenamefont{Kim et~al.}(2005)\citenamefont{Kim, Sachrajda, and
  Sharpe}}]{Kim:2005gf}
\bibinfo{author}{\bibfnamefont{C.}~\bibnamefont{Kim}},
  \bibinfo{author}{\bibfnamefont{C.}~\bibnamefont{Sachrajda}},
  \bibnamefont{and} \bibinfo{author}{\bibfnamefont{S.~R.}
  \bibnamefont{Sharpe}}, \bibinfo{journal}{Nucl.Phys.}
  \textbf{\bibinfo{volume}{B727}}, \bibinfo{pages}{218} (\bibinfo{year}{2005}),
  \eprint{hep-lat/0507006}.

\bibitem[{\citenamefont{Christ et~al.}(2005)\citenamefont{Christ, Kim, and
  Yamazaki}}]{Christ:2005gi}
\bibinfo{author}{\bibfnamefont{N.~H.} \bibnamefont{Christ}},
  \bibinfo{author}{\bibfnamefont{C.}~\bibnamefont{Kim}}, \bibnamefont{and}
  \bibinfo{author}{\bibfnamefont{T.}~\bibnamefont{Yamazaki}},
  \bibinfo{journal}{Phys.Rev.} \textbf{\bibinfo{volume}{D72}},
  \bibinfo{pages}{114506} (\bibinfo{year}{2005}), \eprint{hep-lat/0507009}.

\bibitem[{\citenamefont{Meyer}(2012)}]{Meyer:2012wk}
\bibinfo{author}{\bibfnamefont{H.~B.} \bibnamefont{Meyer}}
  (\bibinfo{year}{2012}), \eprint{1202.6675}.

\bibitem[{\citenamefont{Hansen and Sharpe}(2012)}]{Hansen:2012tf}
\bibinfo{author}{\bibfnamefont{M.~T.} \bibnamefont{Hansen}} \bibnamefont{and}
  \bibinfo{author}{\bibfnamefont{S.~R.} \bibnamefont{Sharpe}},
  \bibinfo{journal}{Phys.Rev.} \textbf{\bibinfo{volume}{D86}},
  \bibinfo{pages}{016007} (\bibinfo{year}{2012}), \eprint{1204.0826}.

\bibitem[{\citenamefont{Agadjanov et~al.}(2014)\citenamefont{Agadjanov,
  Bernard, Mei\ss ner, and Rusetsky}}]{Agadjanov:2014kha}
\bibinfo{author}{\bibfnamefont{A.}~\bibnamefont{Agadjanov}},
  \bibinfo{author}{\bibfnamefont{V.}~\bibnamefont{Bernard}},
  \bibinfo{author}{\bibfnamefont{U.-G.} \bibnamefont{Mei\ss ner}},
  \bibnamefont{and} \bibinfo{author}{\bibfnamefont{A.}~\bibnamefont{Rusetsky}},
  \bibinfo{journal}{Nucl.Phys.} \textbf{\bibinfo{volume}{B886}},
  \bibinfo{pages}{1199} (\bibinfo{year}{2014}), \eprint{1405.3476}.

\bibitem[{\citenamefont{Brice\~no
  et~al.}(2014{\natexlab{b}})\citenamefont{Brice\~no, Hansen, and
  Walker-Loud}}]{Briceno:2014uqa}
\bibinfo{author}{\bibfnamefont{R.~A.} \bibnamefont{Brice\~no}},
  \bibinfo{author}{\bibfnamefont{M.~T.} \bibnamefont{Hansen}},
  \bibnamefont{and}
  \bibinfo{author}{\bibfnamefont{A.}~\bibnamefont{Walker-Loud}}
  (\bibinfo{year}{2014}{\natexlab{b}}), \eprint{1406.5965}.

\bibitem[{\citenamefont{Detmold and Savage}(2004)}]{Detmold:2004qn}
\bibinfo{author}{\bibfnamefont{W.}~\bibnamefont{Detmold}} \bibnamefont{and}
  \bibinfo{author}{\bibfnamefont{M.~J.} \bibnamefont{Savage}},
  \bibinfo{journal}{Nucl.Phys.} \textbf{\bibinfo{volume}{A743}},
  \bibinfo{pages}{170} (\bibinfo{year}{2004}), \eprint{hep-lat/0403005}.

\bibitem[{\citenamefont{Brice\~no and
  Davoudi}(2013{\natexlab{a}})}]{Briceno:2012yi}
\bibinfo{author}{\bibfnamefont{R.~A.} \bibnamefont{Brice\~no}} \bibnamefont{and}
  \bibinfo{author}{\bibfnamefont{Z.}~\bibnamefont{Davoudi}},
  \bibinfo{journal}{Phys. Rev. D. 88,} \textbf{\bibinfo{volume}{094507}},
  \bibinfo{pages}{094507} (\bibinfo{year}{2013}{\natexlab{a}}),
  \eprint{1204.1110}.

\bibitem[{\citenamefont{Bernard et~al.}(2012)\citenamefont{Bernard, Hoja,
  Mei\ss ner, and Rusetsky}}]{Bernard:2012bi}
\bibinfo{author}{\bibfnamefont{V.}~\bibnamefont{Bernard}},
  \bibinfo{author}{\bibfnamefont{D.}~\bibnamefont{Hoja}},
  \bibinfo{author}{\bibfnamefont{U.-G.} \bibnamefont{Mei\ss ner}},
  \bibnamefont{and} \bibinfo{author}{\bibfnamefont{A.}~\bibnamefont{Rusetsky}},
  \bibinfo{journal}{JHEP} \textbf{\bibinfo{volume}{1209}}, \bibinfo{pages}{023}
  (\bibinfo{year}{2012}), \eprint{1205.4642}.

\bibitem[{\citenamefont{Detmold and Flynn}(2014)}]{Detmold:2014fpa}
\bibinfo{author}{\bibfnamefont{W.}~\bibnamefont{Detmold}} \bibnamefont{and}
  \bibinfo{author}{\bibfnamefont{M.}~\bibnamefont{Flynn}}
  (\bibinfo{year}{2014}), \eprint{1412.3895}.

\bibitem[{\citenamefont{Christ}(2010)}]{Christ:2010gi}
\bibinfo{author}{\bibfnamefont{N.~H.} \bibnamefont{Christ}}
  (\bibinfo{collaboration}{RBC Collaboration, UKQCD Collaboration})
  (\bibinfo{year}{2010}), \eprint{1012.6034}.

\bibitem[{\citenamefont{Duncan et~al.}(1996)\citenamefont{Duncan, Eichten, and
  Thacker}}]{Duncan:1996xy}
\bibinfo{author}{\bibfnamefont{A.}~\bibnamefont{Duncan}},
  \bibinfo{author}{\bibfnamefont{E.}~\bibnamefont{Eichten}}, \bibnamefont{and}
  \bibinfo{author}{\bibfnamefont{H.}~\bibnamefont{Thacker}},
  \bibinfo{journal}{Phys.Rev.Lett.} \textbf{\bibinfo{volume}{76}},
  \bibinfo{pages}{3894} (\bibinfo{year}{1996}), \eprint{hep-lat/9602005}.

\bibitem[{\citenamefont{Hayakawa and Uno}(2008)}]{Hayakawa:2008an}
\bibinfo{author}{\bibfnamefont{M.}~\bibnamefont{Hayakawa}} \bibnamefont{and}
  \bibinfo{author}{\bibfnamefont{S.}~\bibnamefont{Uno}},
  \bibinfo{journal}{Prog.Theor.Phys.} \textbf{\bibinfo{volume}{120}},
  \bibinfo{pages}{413} (\bibinfo{year}{2008}), \eprint{0804.2044}.

\bibitem[{\citenamefont{Davoudi and Savage}(2014)}]{Davoudi:2014qua}
\bibinfo{author}{\bibfnamefont{Z.}~\bibnamefont{Davoudi}} \bibnamefont{and}
  \bibinfo{author}{\bibfnamefont{M.~J.} \bibnamefont{Savage}},
  \bibinfo{journal}{Phys.Rev.} \textbf{\bibinfo{volume}{D90}},
  \bibinfo{pages}{054503} (\bibinfo{year}{2014}), \eprint{1402.6741}.

\bibitem[{\citenamefont{Borsanyi et~al.}(2014)\citenamefont{Borsanyi, Durr,
  Fodor, Hoelbling, Katz et~al.}}]{Borsanyi:2014jba}
\bibinfo{author}{\bibfnamefont{S.}~\bibnamefont{Borsanyi}},
  \bibinfo{author}{\bibfnamefont{S.}~\bibnamefont{Durr}},
  \bibinfo{author}{\bibfnamefont{Z.}~\bibnamefont{Fodor}},
  \bibinfo{author}{\bibfnamefont{C.}~\bibnamefont{Hoelbling}},
  \bibinfo{author}{\bibfnamefont{S.}~\bibnamefont{Katz}}, \bibnamefont{et~al.}
  (\bibinfo{year}{2014}), \eprint{1406.4088}.

\bibitem[{\citenamefont{Carrasco et~al.}(2015)\citenamefont{Carrasco, Lubicz,
  Martinelli, Sachrajda, Tantalo et~al.}}]{Carrasco:2015xwa}
\bibinfo{author}{\bibfnamefont{N.}~\bibnamefont{Carrasco}},
  \bibinfo{author}{\bibfnamefont{V.}~\bibnamefont{Lubicz}},
  \bibinfo{author}{\bibfnamefont{G.}~\bibnamefont{Martinelli}},
  \bibinfo{author}{\bibfnamefont{C.}~\bibnamefont{Sachrajda}},
  \bibinfo{author}{\bibfnamefont{N.}~\bibnamefont{Tantalo}},
  \bibnamefont{et~al.} (\bibinfo{year}{2015}), \eprint{1502.00257}.

\bibitem[{\citenamefont{Beane and Savage}(2014)}]{Beane:2014qha}
\bibinfo{author}{\bibfnamefont{S.~R.} \bibnamefont{Beane}} \bibnamefont{and}
  \bibinfo{author}{\bibfnamefont{M.~J.} \bibnamefont{Savage}},
  \bibinfo{journal}{Phys.Rev.} \textbf{\bibinfo{volume}{D90}},
  \bibinfo{pages}{074511} (\bibinfo{year}{2014}), \eprint{1407.4846}.

\bibitem[{\citenamefont{Wasem}(2012)}]{Wasem:2011zz}
\bibinfo{author}{\bibfnamefont{J.}~\bibnamefont{Wasem}},
  \bibinfo{journal}{Phys.Rev.} \textbf{\bibinfo{volume}{C85}},
  \bibinfo{pages}{022501} (\bibinfo{year}{2012}), \eprint{1108.1151}.

\bibitem[{\citenamefont{Rupak and Lee}(2013)}]{Rupak:2013aue}
\bibinfo{author}{\bibfnamefont{G.}~\bibnamefont{Rupak}} \bibnamefont{and}
  \bibinfo{author}{\bibfnamefont{D.}~\bibnamefont{Lee}},
  \bibinfo{journal}{Phys.Rev.Lett.} \textbf{\bibinfo{volume}{111}},
  \bibinfo{pages}{032502} (\bibinfo{year}{2013}), \eprint{1302.4158}.

\bibitem[{\citenamefont{L\"uscher}(1986)}]{Luscher:1986pf}
\bibinfo{author}{\bibfnamefont{M.}~\bibnamefont{L\"uscher}},
  \bibinfo{journal}{Commun.Math.Phys.} \textbf{\bibinfo{volume}{105}},
  \bibinfo{pages}{153} (\bibinfo{year}{1986}).

\bibitem[{\citenamefont{Luscher}(1991)}]{Luscher:1990ux}
\bibinfo{author}{\bibfnamefont{M.}~\bibnamefont{L\"uscher}},
  \bibinfo{journal}{Nucl.Phys.} \textbf{\bibinfo{volume}{B354}},
  \bibinfo{pages}{531} (\bibinfo{year}{1991}).

\bibitem[{\citenamefont{Rummukainen and Gottlieb}(1995)}]{Rummukainen:1995vs}
\bibinfo{author}{\bibfnamefont{K.}~\bibnamefont{Rummukainen}} \bibnamefont{and}
  \bibinfo{author}{\bibfnamefont{S.~A.} \bibnamefont{Gottlieb}},
  \bibinfo{journal}{Nucl. Phys.} \textbf{\bibinfo{volume}{B450}},
  \bibinfo{pages}{397} (\bibinfo{year}{1995}), \eprint{hep-lat/9503028}.

\bibitem[{\citenamefont{Brice\~no}(2014{\natexlab{b}})}]{Briceno:2014oea}
\bibinfo{author}{\bibfnamefont{R.~A.} \bibnamefont{Brice\~no}},
  \bibinfo{journal}{Phys.Rev.} \textbf{\bibinfo{volume}{D89}},
  \bibinfo{pages}{074507} (\bibinfo{year}{2014}{\natexlab{b}}),
  \eprint{1401.3312}.

\bibitem[{\citenamefont{Polejaeva and Rusetsky}(2012)}]{Polejaeva:2012ut}
\bibinfo{author}{\bibfnamefont{K.}~\bibnamefont{Polejaeva}} \bibnamefont{and}
  \bibinfo{author}{\bibfnamefont{A.}~\bibnamefont{Rusetsky}},
  \bibinfo{journal}{Eur.Phys.J.} \textbf{\bibinfo{volume}{A48}},
  \bibinfo{pages}{67} (\bibinfo{year}{2012}), \eprint{1203.1241}.

\bibitem[{\citenamefont{Brice\~no and
  Davoudi}(2013{\natexlab{b}})}]{Briceno:2012rv}
\bibinfo{author}{\bibfnamefont{R.~A.} \bibnamefont{Brice\~no}} \bibnamefont{and}
  \bibinfo{author}{\bibfnamefont{Z.}~\bibnamefont{Davoudi}},
  \bibinfo{journal}{Phys.Rev.} \textbf{\bibinfo{volume}{D87}},
  \bibinfo{pages}{094507} (\bibinfo{year}{2013}{\natexlab{b}}),
  \eprint{1212.3398}.

\bibitem[{\citenamefont{Hansen and Sharpe}(2014)}]{Hansen:2014eka}
\bibinfo{author}{\bibfnamefont{M.~T.} \bibnamefont{Hansen}} \bibnamefont{and}
  \bibinfo{author}{\bibfnamefont{S.~R.} \bibnamefont{Sharpe}},
  \bibinfo{journal}{Phys.Rev.} \textbf{\bibinfo{volume}{D90}},
  \bibinfo{pages}{116003} (\bibinfo{year}{2014}), \eprint{1408.5933}.

\bibitem[{\citenamefont{McNeile and Michael}(2006)}]{McNeile:2006qy}
\bibinfo{author}{\bibfnamefont{C.}~\bibnamefont{McNeile}} \bibnamefont{and}
  \bibinfo{author}{\bibfnamefont{C.}~\bibnamefont{Michael}}
  (\bibinfo{collaboration}{UKQCD Collaboration}), \bibinfo{journal}{Phys.Lett.}
  \textbf{\bibinfo{volume}{B642}}, \bibinfo{pages}{244} (\bibinfo{year}{2006}),
  \eprint{hep-lat/0607032}.

\bibitem[{\citenamefont{Mastropas and Richards}(2014)}]{Mastropas:2014fsa}
\bibinfo{author}{\bibfnamefont{E.~V.} \bibnamefont{Mastropas}}
  \bibnamefont{and} \bibinfo{author}{\bibfnamefont{D.~G.}
  \bibnamefont{Richards}} (\bibinfo{collaboration}{Hadron Spectrum}),
  \bibinfo{journal}{Phys.Rev.} \textbf{\bibinfo{volume}{D90}},
  \bibinfo{pages}{014511} (\bibinfo{year}{2014}), \eprint{1403.5575}.

\bibitem[{\citenamefont{Li and Liu}(2013)}]{Li:2012bi}
\bibinfo{author}{\bibfnamefont{N.}~\bibnamefont{Li}} \bibnamefont{and}
  \bibinfo{author}{\bibfnamefont{C.}~\bibnamefont{Liu}},
  \bibinfo{journal}{Phys.Rev.} \textbf{\bibinfo{volume}{D87}},
  \bibinfo{pages}{014502} (\bibinfo{year}{2013}), \eprint{1209.2201}.

\bibitem[{\citenamefont{Li et~al.}(2014)\citenamefont{Li, Li, and
  Liu}}]{Li:2014wga}
\bibinfo{author}{\bibfnamefont{N.}~\bibnamefont{Li}},
  \bibinfo{author}{\bibfnamefont{S.-Y.} \bibnamefont{Li}}, \bibnamefont{and}
  \bibinfo{author}{\bibfnamefont{C.}~\bibnamefont{Liu}},
  \bibinfo{journal}{Phys.Rev.} \textbf{\bibinfo{volume}{D90}},
  \bibinfo{pages}{034509} (\bibinfo{year}{2014}), \eprint{1401.5569}.

\bibitem[{\citenamefont{Dudek et~al.}(2014)\citenamefont{Dudek, Edwards,
  Thomas, and Wilson}}]{Dudek:2014qha}
\bibinfo{author}{\bibfnamefont{J.~J.} \bibnamefont{Dudek}},
  \bibinfo{author}{\bibfnamefont{R.~G.} \bibnamefont{Edwards}},
  \bibinfo{author}{\bibfnamefont{C.~E.} \bibnamefont{Thomas}},
  \bibnamefont{and} \bibinfo{author}{\bibfnamefont{D.~J.} \bibnamefont{Wilson}}
  (\bibinfo{year}{2014}), \eprint{1406.4158}.

\bibitem[{\citenamefont{Wilson et~al.}(2014)\citenamefont{Wilson, Dudek,
  Edwards, and Thomas}}]{Wilson:2014cna}
\bibinfo{author}{\bibfnamefont{D.~J.} \bibnamefont{Wilson}},
  \bibinfo{author}{\bibfnamefont{J.~J.} \bibnamefont{Dudek}},
  \bibinfo{author}{\bibfnamefont{R.~G.} \bibnamefont{Edwards}},
  \bibnamefont{and} \bibinfo{author}{\bibfnamefont{C.~E.} \bibnamefont{Thomas}}
  (\bibinfo{year}{2014}), \eprint{1411.2004}.

\bibitem[{\citenamefont{Moore and Fleming}(2006)}]{Moore:2005dw}
\bibinfo{author}{\bibfnamefont{D.~C.} \bibnamefont{Moore}} \bibnamefont{and}
  \bibinfo{author}{\bibfnamefont{G.~T.} \bibnamefont{Fleming}},
  \bibinfo{journal}{Phys.Rev.} \textbf{\bibinfo{volume}{D73}},
  \bibinfo{pages}{014504} (\bibinfo{year}{2006}), \eprint{hep-lat/0507018}.

\bibitem[{\citenamefont{Dudek et~al.}(2013)\citenamefont{Dudek, Edwards, and
  Thomas}}]{Dudek:2012xn}
\bibinfo{author}{\bibfnamefont{J.~J.} \bibnamefont{Dudek}},
  \bibinfo{author}{\bibfnamefont{R.~G.} \bibnamefont{Edwards}},
  \bibnamefont{and} \bibinfo{author}{\bibfnamefont{C.~E.}
  \bibnamefont{Thomas}}, \bibinfo{journal}{Phys.Rev.}
  \textbf{\bibinfo{volume}{D87}}, \bibinfo{pages}{034505}
  (\bibinfo{year}{2013}), \eprint{1212.0830}.

\bibitem[{\citenamefont{Bedaque}(2004)}]{Bedaque:2004kc}
\bibinfo{author}{\bibfnamefont{P.~F.} \bibnamefont{Bedaque}},
  \bibinfo{journal}{Phys.Lett.} \textbf{\bibinfo{volume}{B593}},
  \bibinfo{pages}{82} (\bibinfo{year}{2004}), \eprint{nucl-th/0402051}.

\bibitem[{\citenamefont{Brice\~no
  et~al.}(2014{\natexlab{c}})\citenamefont{Brice\~no, Davoudi, Luu, and
  Savage}}]{Briceno:2013hya}
\bibinfo{author}{\bibfnamefont{R.~A.} \bibnamefont{Brice\~no}},
  \bibinfo{author}{\bibfnamefont{Z.}~\bibnamefont{Davoudi}},
  \bibinfo{author}{\bibfnamefont{T.~C.} \bibnamefont{Luu}}, \bibnamefont{and}
  \bibinfo{author}{\bibfnamefont{M.~J.} \bibnamefont{Savage}},
  \bibinfo{journal}{Phys.Rev.} \textbf{\bibinfo{volume}{D89}},
  \bibinfo{pages}{074509} (\bibinfo{year}{2014}{\natexlab{c}}),
  \eprint{1311.7686}.

\end{thebibliography}
\end{document}